\documentclass[a4paper,UKenglish]{lipics-v2016}
 
\usepackage{microtype}

\title{On Reversible Transducers
}
\titlerunning{On Reversible Transducers} 

\author[1]{Luc Dartois}
\author[2]{Paulin Fournier}
\author[1]{Isma\"{e}l Jecker}
\author[1,2]{Nathan Lhote}
\affil[1]{Universit\'{e} Libre de Bruxelles\\
  \texttt{\{ldartois, ijecker\}@ulb.ac.be}}
\affil[2]{Universit\'e de Bordeaux, LaBRI\\
  \texttt{\{paulin.fournier,nlhote\}@labri.fr}}
\authorrunning{L. Dartois and P. Fournier and I. Jecker and N. Lhote} 

\Copyright{Luc Dartois and Paulin Fournier and Isma\"{e}l Jecker and Nathan Lhote}


\EventEditors{John Q. Open and Joan R. Acces}
\EventNoEds{2}
\EventLongTitle{42nd Conference on Very Important Topics (CVIT 2016)}
\EventShortTitle{CVIT 2016}
\EventAcronym{CVIT}
\EventYear{2016}
\EventDate{December 24--27, 2016}
\EventLocation{Little Whinging, United Kingdom}
\EventLogo{}
\SeriesVolume{42}
\ArticleNo{23}

\usepackage{mathabx}

\usepackage{comment}
\usepackage{mathtools}
\usepackage{amsthm}
\usepackage{amssymb}
\usepackage{cite}
\usepackage{tikz}
\usepackage{graphicx}
\usepackage[basic,classfont=caps]{complexity}
\usepackage[T1]{fontenc}
\usepackage{enumerate}
\usepackage{paralist}
\usepackage{algorithm}
\usepackage{algorithmic}
\usepackage{xspace}
\usepackage{accents}
\usepackage{stmaryrd}
\usepackage{enumitem}
\usepackage{todonotes}
\usepackage{microtype}
\usepackage{xxcolor}
\usepackage{thm-restate}

\usetikzlibrary{chains,matrix,scopes,decorations.shapes,arrows,shapes}

\usepackage{wrapfig}
\usepackage{scrextend}

\newcommand{\myparagraph}[1]{\par\smallskip\noindent\textbf{#1.}}

\makeatletter
\let\epsilon\varepsilon
\let\phi\varphi
\let\emptyset\varnothing
\let\rho\varrho
\makeatother

\renewclass{\P}{PTime}
\renewclass{\DTIME}{DTime}
\renewclass{\DSPACE}{DSpace}
\renewclass{\EXP}{ExpTime}
\newclass{\TWOEXP}{2ExpTime}
\renewclass{\EXPSPACE}{ExpSpace}
\newclass{\ACK}{Ack}
\newclass{\FDTIME}{FTtime}
\renewclass{\AP}{APTime}
\renewclass{\PSPACE}{PSpace}

\usetikzlibrary{positioning,fit,arrows,automata,calc,shapes,decorations.pathmorphing} 
\tikzset{->,>=stealth',
shorten >=1pt,shorten <=1pt,
auto,node distance=1.5cm,
every loop/.style={looseness=6},
initial text={},
every state/.style={inner sep=0.2mm, minimum size=0.5cm},
el/.style={font=\scriptsize},
inner sep=1mm,
loopright/.style={loop,looseness=6,out=30, in=-30},
loopleft/.style={loop,looseness=6,out=210, in=150},
loopabove/.style={loop,looseness=6,out=120, in=60},
loopbelow/.style={loop,looseness=6,out=300, in=240},
ve/.style={rectangle,draw,inner sep=1mm},
va/.style={circle,draw,inner sep=0.6mm},
}

\newcommand{\var}{\mathcal{X}}
\newcommand{\varO}{O}
\newcommand{\aut}{\mathcal{A}}
\newcommand{\odet}{\mathcal{D}}
\newcommand{\unif}{\mathcal{U}}
\newcommand{\tra}{\mathcal{T}}
\newcommand{\mult}[1][]{\mathcal{M}_{#1}}
\newcommand{\idsub}[1]{\textsf{Id}_{#1}}
\newcommand{\empsub}{\sigma_\epsilon}
\providecommand\lang{}
\renewcommand{\lang}{\mathcal{L}}
\newcommand{\trans}{\mathcal{R}}
\newcommand{\sst}{\mathcal{Z}}
\newcommand{\alp}{A}
\newcommand{\alpo}{B}

\newcommand{\sub}[2]{\mathcal{S}_{#1,#2}}
\newcommand{\tFA}{2FA}
\newcommand{\oFA}{1FA}

\newcommand{\tFT}{2FT}
\newcommand{\oFT}{1FT}
\newcommand{\SST}{SST}
\newcommand{\vdashv}[1]{#1_{\vdash \dashv}}

\newcommand{\preclex}{\prec_{\textsf{lex}}}
\newcommand{\precsli}{\prec_{\textsf{sl}}}

\newcommand{\state}{\mathcal{F}}

\newcommand{\delLRb}[1]{{\alpha}^{+}_{#1}}

\newcommand{\delRLb}[1]{{\beta}^{-}_{#1}}
\newcommand{\delRRAb}[1]{{\beta}^{+}_{#1}}
\newcommand{\delRRBb}[1]{{\beta}^{+}_{#1}}

\newcommand{\lbeh}[1]{\textsf{B}_{\ell}(#1)}

\newcommand{\llbeh}[1]{\textsf{R}_{\ell \ell}(#1)}

\newcommand{\predfin}[1]{\textsf{F}_{\ell \dashv}(#1)}

\newcommand{\projrun}[2]{{\pi}_{#1}(#2)}
\newcommand{\sta}[2]{(#1,#2)}
\newcommand{\staf}[2]{#1 \ \ #2}
\newcommand{\succmax}[1]{#1_{\max}}
\newcommand{\succmin}[1]{#1_{\min}}
\newcommand{\dead}[1]{\overline{#1}}
\newcommand{\live}[1]{\underline{#1}}
\newcommand{\longrun}[1]{\lambda_{#1}}
\newcommand{\lonesuc}[1]{{#1}_{0}}

\newcommand{\nathan}[1]{\todo[color=green!50!blue!30,inline]{N: #1}}
\newcommand{\ismael}[1]{\todo[color=yellow!30,inline]{I: #1}}
\newcommand{\luc}[1]{\todo[color=red!30,inline]{L: #1}}

\newcommand{\ov}[1]{\overline{#1}}
\newcommand{\ud}[1]{\underline{#1}}

\newcommand{\pair}[2]{{(#1,#2)}}
\newcommand{\tr}[3]{\displaystyle \left(#1,#2,#3\right)}
\newcommand{\trs}[4]{\tr{\pair{#1}{#2}}{a}{\pair{#3}{#4}}}
\newcommand{\fua}{\textbf{(fua)}}
\newcommand{\fuw}{\textbf{(fuw)}}
\newcommand{\flw}{\textbf{(flw)}}
\newcommand{\fla}{\textbf{(fla)}}
\newcommand{\buw}{\textbf{(buw)}}
\newcommand{\bua}{\textbf{(bua)}}
\newcommand{\bla}{\textbf{(bla)}}
\newcommand{\blw}{\textbf{(blw)}}
\newcommand{\fualw}{\textbf{(fualw)}}
\newcommand{\fuwla}{\textbf{(fuwla)}}
\newcommand{\bulw}{\textbf{(bulw)}}
\newcommand{\bula}{\textbf{(bula)}}

\begin{document}

\maketitle

\begin{abstract}
Deterministic two-way transducers define the robust class of regular functions which is, among other good properties, closed under composition.
However, the best known algorithms for composing two-way transducers cause a double exponential blow-up in the size of the inputs.
In this paper, we introduce a class of transducers for which the composition has polynomial complexity. It is the class of reversible transducers, for which the computation steps can be reversed deterministically.
While in the one-way setting this class is not very expressive, we prove that any two-way transducer can be made reversible through a single exponential blow-up.
As a consequence, we prove that the composition of two-way transducers can be done with a single exponential blow-up in the number of states.

A uniformization of a relation is a function with the same domain and which is included in the original relation.
Our main result actually states that we can uniformize any non-deterministic two-way transducer by a reversible transducer with a single exponential blow-up, improving the known result by de Souza which has a quadruple exponential complexity.
As a side result, our construction also gives a quadratic transformation from copyless streaming string transducers to two-way transducers, improving the exponential previous bound.
\end{abstract}
 
 \section{Introduction}\label{Section-Intro}

\myparagraph{Automata and transducers}
Automata theory is a prominent domain of theoretical computer science, initiated in the 60s~\cite{Buchi60} and still very active nowadays.
Many extensions of finite automata have been studied such as automata over more complex structures (infinite words, trees, \emph{etc}) or transducers which can be seen as automata with an additional write-only output tape and which will be the focus of our study in the remainder of this article.

Transducers have been studied for almost as long as automata~\cite{AHU69} and important results have been obtained, however the theory of transducers is not as advanced as automata theory. One of the reasons for this is that many descriptions which are equivalent for automata become different in expressiveness in the case of transducers.
For instance, deterministic and non-deterministic automata recognize the same class of languages, the \emph{regular languages}.
However this is not the case for transducers since in particular a deterministic transducer must realize a function while a non-deterministic one may realize a relation.
Similarly, by allowing the reading head to move left and right, one gets a two-way model of automata and it is known that two-way automata are as expressive as one-way automata~\cite{She59}. However two-way transducers can model relations and functions that are unobtainable in the one-way case, such as the function \textsf{mirror} which reverses its input.
Recently, two-way transducers were also proven to be equivalent to the one-way deterministic model of \emph{streaming string transducers}~\cite{AC10}, which can be thought of as transducers with write-only registers.

\myparagraph{Reversible transducers}
A transition system is called \emph{reversible} when for every input, the directed graph of configurations is composed of nodes of in-degree and out-degree at most one.
This property is stronger than the more studied notion of determinism since it allows to navigate back and forth between the steps of a computation.
In this article, we study the class of transducers that are simultaneously deterministic and co-deterministic, \emph{i.e.} reversible.
The main motivation for the definition of this class is its good properties with respect to composition.
When we consider one-way transducers, runs only go forward  and thus determinism gives good properties for composition: the next step of a run is computed in constant time. 
However, when considering composition of two-way transducers, the second machine can move to the left, which corresponds to rewind the run of the first machine.
Then the stronger property of reversibility allows for this back and forth navigation over runs of transducers, and we recover the property of reaching the next (or previous here) step of a computation in a constant time.
This leads to recover the polynomial state complexity of composition that exists for deterministic one-way transducers.
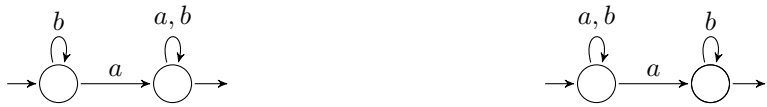
\begin{figure}
\begin{subfigure}{.5\textwidth}
\begin{tikzpicture}
\node[state,initial] (i) {};
\node[state,right of=i] (st) {};
\node[right=0.5 of st] (inv) {};
\path[->] (i) edge [loop above] node {$b$}();
\path[->] (i) edge node {$a$} (st);
\path[->] (st) edge [loop above] node {$a,b$}();
\path[->] (st) edge node {} (inv);

\end{tikzpicture}

 \end{subfigure}
\begin{subfigure}{.5\textwidth}
\begin{tikzpicture}
\node[state,initial] (i) {};
\node[state,right of=i] (st) {};
\node[state,right of=i] (st) {};
\node[right=0.5 of st] (inv) {};
\path[->] (i) edge [loop above] node {$a,b$}();
\path[->] (i) edge node {$a$} (st);
\path[->] (st) edge [loop above] node {$b$}();
\path[->] (st) edge node {} (inv);
\end{tikzpicture}
 \end{subfigure}
\caption{The language $A^*aA^*$ can be recognized by a deterministic (left) or co-deterministic (right) automaton, but not by a reversible one.}
\label{fig::det-cod}
\end{figure}

Let us now discuss the expressiveness of reversible transducers.
Regarding automata in the one-way case, it is well-known that any regular language can be recognized by a deterministic one-way automaton and symmetrically by a co-deterministic one-way automaton, since the mirror of a regular language is still regular.
However, the class of one-way reversible automata is very restrictive (see Figure~\ref{fig::det-cod} for an example or~\cite{Pin92} for a study of its expressive power, where they are called bideterministic).
It turns out however, that if we allow bidirectionality then any regular language can be recognized by a reversible automaton. In fact, a two-way reversible automaton can be constructed from either a one-way or two-way automaton using only a linear number of states (see \cite{KondacsW97} and \cite{KuncO13}, respectively).
We prove, as a consequence of our main theorem, that reversible transducer are as expressive as functional two-way transducers, and exactly capture the class of regular functions.
As states earlier, regular functions are also characterized by streaming string transducers (\SST). As a byproduct, we also give a quadratic construction from copyless \SST to reversible transducers, improving results from~\cite{AFT12,DJR16}.

\myparagraph{Synthesis problem and uniformization of transducers}
In the bigger picture of verification, two-way transducers can be used to model transformations of programs or non-reactive systems.
If we consider the synthesis problem, where the specification is given as a relation of admissible input-output pairs,
an implementation is then given as a function, with the same domain, relating a unique output to a given input.
The uniformization problem asks if given a relation, we can extract a function that has the same domain, and is included in the relation.
We argue that the synthesis problem can be instantiated in the setting of transformations as the problem of uniformization of a non-deterministic two-way transducer by a functional transducer.
Our main result states that we can uniformize any non-deterministic two-way transducer by a reversible transducer with a single exponential blow-up.

\myparagraph{Related work}
As stated earlier, reversible one-way automata were already considered in~\cite{Pin92}.
Two-way reversible automata were shown to capture the regular languages in~\cite{KondacsW97} by a construction from a one-way deterministic automaton to a two-way reversible automaton with a linear blow-up. 
This construction was extended to two-way automata in~\cite{KuncO13}, still with a linear complexity. 
However, these constructions for automata cannot be simply extended to transducers because more information is needed in order to produce the outputs at the right moment.
To the best of our knowledge, reversible transducers have not been studied yet, however, since we introduce reversible transducers as a tool for the composition of transducers, our work can be linked with the construction of Hopcroft and Ullman that gives the composition of a one-way transducer and a two-way transducer, while preserving determinism. 
Our construction strictly improves theirs, since ours produces, with a polynomial complexity instead of an exponential one, a reversible transducer that can in turn be easily composed.

A procedure for the uniformization of a two-way non-deterministic transducers by a deterministic on has been known since~\cite{DS13}. 
The complexity of this procedure is quadruple exponential, while ours construction is single exponential, and produces a reversible transducer.

\myparagraph{Organization of the paper}
Preliminary definitions are given in the next Section.
In Section~\ref{Section:Results}, we present our main results on composability and expressiveness of reversible transducers.
Section~\ref{Section:proof} is devoted to the main technical construction of the paper.
Connections with streaming string transducers are discussed in Section~\ref{Section:sst} while further works are considered in Section~\ref{Section:Concl}.

 \section{Automata and transducers}\label{Section:prelim}

Given a finite alphabet $\alp$, we denote by $\alp^*$ the set of finite words over $\alp$, and by $\epsilon$ the empty word.
We will denote by $\vdashv{\alp}$ the alphabet $A \uplus \{\vdash, \dashv \}$, where the new symbols $\vdash$ and $\dashv$ are called \emph{endmarkers}.
A \emph{language} over $\alp$ is a subset $\lang$ of $\alp^*$.
Given two finite alphabets $\alp$ and $\alpo$, a \emph{transduction} from $\alp$ to $\alpo$ is a relation $\trans \subseteq \alp^* \times \alpo^*$.

\myparagraph{Automata}
A \emph{two-way finite state automaton} (\tFA) is a tuple $\aut = (\alp, Q, q_I, q_F, \Delta)$, where $\alp$ is a finite alphabet;
$Q$ is a finite set of states partitioned into the set of forward states $Q^{+}$ and the set of backward states $Q^{-}$;
$q_I \in Q^{+}$ is the initial state;
$q_F \in Q^{+}$ is the final state;
$\Delta \subseteq Q \times \vdashv{\alp} \times Q$ is the state transition relation.
By convention, $q_I$ and $q_F$ are the only forward states verifying $(q_I,\vdash,q) \in \Delta$
and $(q,\dashv,q_F) \in \Delta$ for some $q \in Q$.
However, for any backward state $p^{-} \in Q^{-}$, $\Delta$ might contain transitions $(p^{-},\vdash,q)$
and $(q,\dashv,p^{-})$, for some $q \in Q$.
Note that, in our figures, we do not represent explicitly the initial and final states, and use arrows labeled with
the endmarkers to indicate the corresponding transitions.
A configuration $u.p.u'$ of $\aut$ is composed of two words $u,u' \in \vdashv{\alp}^*$ and a state $p \in Q$.
The configuration $u.p.u'$ admits a set of successor configurations, defined as follows.
If $p \in Q^+$, the input head currently reads the first letter of the suffix $u' = a'v'$.
The successor of $u.p.u'$ after a transition $(p,a',q)\in \Delta$ is either $ua'.q.v'$ if $q\in Q^+$, or 
$u.q.u'$ if $q\in Q^-$.
Conversely, if $p \in Q^-$, the input head currently reads the last letter of the prefix $u = v a$.
The successor of $u.p.u'$ after $(p,a',q)\in \Delta$ is $u.q.u'$ if $q\in Q^+$, or $v.q.au'$ if $q\in Q^-$.
For every word $u \in \vdashv{\alp}^*$, a \emph{run} of $\aut$ on $u$ is a sequence of successive configurations
$\rho = u_0.q_0.u_0',\ldots, u_m.q_m.u_m'$ such that for every $0 \leq i \leq m$, $u_iu_{i}' = u$.
The run $\rho$ is called \emph{initial} if it starts in configuration $q_I.u$, \emph{final} if it ends in configuration $u.q_F$,
\emph{accepting} if it is both initial and final, and \emph{end-to-end} if it starts and ends on the boundaries of $u$.
More precisely, it is called
\emph{left-to-right} if $q_0,q_m \in Q^+$ and $u_0 = u_m' = \epsilon$;
\emph{right-to-left} if $q_0,q_m \in Q^-$ and $u_0' = u_m = \epsilon$;
\emph{left-to-left} if $q_0 \in Q^+$, $q_m \in Q^-$ and $u_0 = u_m = \epsilon$;
\emph{right-to-right} if $q_0 \in Q^-$, $q_m \in Q^+$ and $u_0' = u_m' = \epsilon$.
Abusing notations, we also denote by $\Delta$ the extension of the state transition relation to a subset of $Q \times \vdashv{\alp}^* \times Q$
composed of the triples $(p,u,q)$ such that there exists an end-to-end run on $u$ between $p$ and $q$.
For every triple $(p,u,q) \in \Delta$, we say that $q$ is a \emph{$u$-successor} of $p$ and
that $p$ is a \emph{$u$-predecessor} of $q$.
The language $\lang_{\aut}$ \emph{recognized} by $\aut$ is the set of words $u \in \alp^*$ such that $\vdash u \dashv$
admits an accepting run, i.e., $(q_I,\vdash u \dashv,q_F) \in \Delta$.
The automaton $\aut$ is called 
\begin{itemize}
\item
a \emph{one-way finite state automaton} (\oFA{}) if the set $Q^-$ is empty;
\item
\emph{deterministic} if for all $(p,a) \in Q \times \vdashv{\alp}$, there is at most one $q \in Q$ verifying $(p,a,q) \in \Delta$;
\item
\emph{weakly branching} if for all $a \in \alp$ there is at most one state $p \in Q$ and one pair of distinct states $q_1,q_2 \in Q$ such that $(p,a,q_1) \in \Delta$ and $(p,a,q_2) \in \Delta$.
\item \emph{co-deterministic} if for all $(q,a) \in Q \times \vdashv{\alp}$, there is at most one $p \in Q$ verifying $(p,a,q) \in \Delta$;
\item
\emph{reversible} if it is both deterministic and co-deterministic.
\end{itemize}
An automaton with several initial and final states can be simulated by using non-determinism while reading the endmarker $\vdash$
and non-co-determinism while reading the endmarker $\dashv$, hence
requiring a single initial state and a single final state does not restrict the expressiveness of our model.

\myparagraph{Transducers}
A \emph{two-way finite state transducer} is a tuple $\tra = (\alp,\alpo, Q, q_I, q_F, \Delta, \mu)$, where $\alpo$ is a finite alphabet;
$\aut_{\tra} = (\alp, Q, q_I, q_F, \Delta)$ is a \tFA, called the \emph{underlying automaton} of $\tra$;
and $\mu : \Delta \rightarrow \alpo^*$ is the output function.
A run of $\tra$ is a run of its underlying automaton, and the language $\lang_{\tra}$ recognized by $\tra$
is the language $\lang_{\aut_{\tra}} \in \alp^*$ recognized by its underlying automaton.
Given a run $\rho$ of $\tra$, we set $\mu(\rho) \in \alpo^*$ as the concatenation of the images by $\mu$ of the transitions of $\tra$
occurring along $\rho$.
Note that in the deterministic (or co-deterministic) case we are able to extend $\mu$ to end-to-end runs since in this case we can firmly associate an end-to-end run to a unique sequence of transitions $(p,u,q)$.
The transduction $\trans_{\tra} \subseteq \alp^* \times \alpo^*$ defined by $\tra$ is the set of pairs $(u,v)$
such that $u \in \lang_{\tra}$ and $\mu(\rho) = v$ for an accepting run $\rho$ of $\aut_{\tra}$ on $\vdash u \dashv$.
Two transducers are called \emph{equivalent} if they define the same transduction.
A transducer $\tra$ is respectively called one-way, deterministic, weakly branching, co-deterministic or reversible, if its underlying automaton has the corresponding property.

\begin{figure}
\begin{subfigure}{.5\textwidth}
\begin{tikzpicture}[scale=0.7]
\node[circle,draw,inner sep=4] (n0) at (0,0) {$0$} ;
\node[circle,draw,inner sep=4] (n1) at (2,0) {$1$} ;
\node[circle,draw,inner sep=4] (n2) at (4,0) {$2$};
\node (s) at (0,-2.5){};
\draw[>=stealth]  (-1,0) -> node[midway,above] {$\vdash$} (n0);
\draw[>=stealth]  (n2) -> node[midway,above] {$\dashv$}  (5,0);
\draw (n0) edge  node[midway,above] {$a$} (n1);
\draw (n1) edge  node[midway,above] {$a$} (n2);
\draw (n1) edge[bend left]  node[midway,below] {$b$} (n0);
\draw (n0) edge[loop above] node[midway, above]{$b$} (n0);
\draw (n2) edge[loop above] node[midway, above]{$a,b$} (n2);
\end{tikzpicture}
 \caption{A deterministic \oFA{} $\aut_1$}
\label{A1}
 \end{subfigure}
\begin{subfigure}{.5\textwidth}
\begin{tikzpicture}[scale=0.7]

\providecommand\XA{}
\renewcommand{\XA}{0.23}

\providecommand\YA{}
\renewcommand{\YA}{0.17}

\providecommand\XB{}
\renewcommand{\XB}{0.}

\providecommand\YB{}
\renewcommand{\YB}{0.23}

\node[circle,draw,inner sep=4] (n0) at (0,0) {$0$} ;
\node[] (n) at (\XA,\YA) {\tiny $+$} ;
\node[circle,draw,inner sep=4] (n1) at (2,0) {$1$} ;
\node[] (n) at ( 2 +\XA, 0 +\YA) {\scriptsize $+$} ;
\node[circle,draw,inner sep=4] (n2) at (4,0) {$1$};
\node[] (n) at ( 4 +\XA, 0 +\YA) {\scriptsize $-$} ;
\node[] (n) at (4 +\XB,\YB) {\tiny $\sim$} ;
\node[circle,draw,inner sep=4] (n1b) at (2,-2) {$1$} ;
\node[] (n) at ( 2 +\XA, -2 +\YA) {\scriptsize $-$} ;
\node[circle,draw,inner sep=4] (n2b) at (4,-2) {$1$};
\node[] (n) at ( 4 +\XA, -2 +\YA) {\scriptsize $+$} ;
\node[] (n) at (4 +\XB,-2 +\YB) {\tiny $\sim$} ;
\node[circle,draw,inner sep=4] (n3) at (6,0) {$0$};
\node[] (n) at ( 6 +\XA, 0 +\YA) {\scriptsize $-$} ;
\node[] (n) at (6 +\XB,\YB) {\tiny $\sim$} ;
\node[circle,draw,inner sep=4] (n4) at (8,0) {$2$};
\node[] (n) at ( 8 +\XA, 0 +\YA) {\scriptsize $+$} ;
\node (s) at (0,-2.5){};
\draw[>=stealth]  (-1,0) -> node[midway,above] {$\vdash$} (n0);
\draw[>=stealth]  (n4) -> node[midway,above] {$\dashv$}  (9,0);
\draw (n0) edge  node[midway,above] {$a$} (n1);
\draw (n1) edge  node[midway,right] {$b$} (n1b);
\draw (n1b) edge  node[midway] {$a$} (n0);
\draw (n0) edge[loop above] node[midway, above]{$b$} (n0);
\draw (n1) edge  node[midway,above] {$a$} (n2);
\draw (n2) edge  node[midway,above] {$a$} (n3);
\draw (n2b) edge  node[midway,left] {$b$} (n2);
\draw (n3) edge  node[midway] {$a$} (n2b);
\draw (n3) edge  node[midway,above] {$\vdash$} (n4);
\draw (n3) edge[loop above] node[midway, above]{$b$} (n3);
\draw (n4) edge[loop above] node[midway, above]{$a,b$} (n4);
\end{tikzpicture}
 \caption{A reversible \tFA{} $\aut_2$}
\label{A2}
 \end{subfigure}
\caption{Two automata recognizing the same language.}
\label{fig::2aut}
\end{figure}
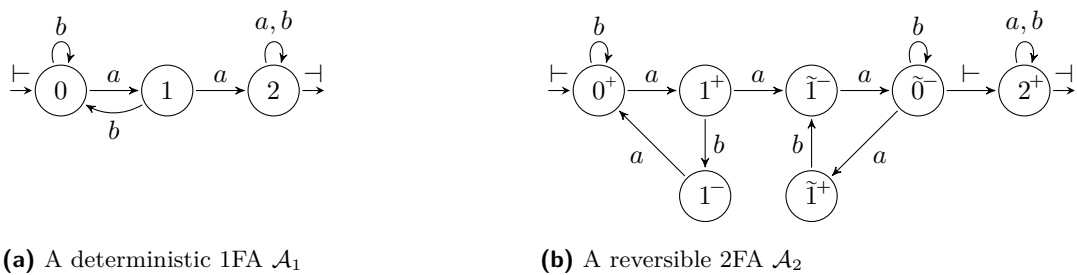

\myparagraph{Examples}
Let us consider the language $\mathcal{L}_{aa} \subseteq \{a,b\}^*$ composed of the  words
that contain two $a$ symbols in a row. This language is recognized by the deterministic one-way automaton $\aut_1$,
represented in Figure \ref{A1}, and by the reversible two-way automaton $\aut_2$, represented in Figure~\ref{A2}.
However, it is not recognizable by a one-way reversible automaton.
Let us analyze the behavior of $\aut_2$ to see how moving back an forth through the input allows it to recognize
$\mathcal{L}_{aa}$ in a reversible manner. 
First, $\aut_2$ uses an intermediate step to go from $1^+$ back to $0^+$ when reading a
$b$, to avoid creating non-co-determinism. Second, once $\aut_2$ reads two consecutive $a$ symbols, it does not go
directly in the final state looping on every input, since this would generate non-co-determinism.
Instead, $\aut_2$ goes in an inverse copy of the first three states, where it rewind its run until the left endmarker.
It is then free to go in the looping accepting state.

 \section{Results on Reversible Transducers}\label{Section:Results}
In this section, we present the main results of our paper.
In Subsection~\ref{Subs:Compo}, we show the polynomial composition of reversible transductions.
In the following, we give expressiveness results of the class of reversible transducers, relying on this composition as well as the construction presented in Section~\ref{Section:proof}. 
\subsection{Composition of reversible transducers}\label{Subs:Compo}
The nicest feature of reversible transducers has to be the low complexity (and simplicity) of their composition. Indeed the composition of two such transducers is polynomial in the number of states of the inputs, and the construction itself is quite simple. 
This is due to the fact that the difficult part in the composition of transducers is to be able to navigate the run easily. In the one-way case, the composition is easy since runs can only move forward.
In the two-way case, one needs to advance in the run, but also rewind it. Since the former is made easy by the determinism, and the latter is symmetrically handled by the co-deterministim, composition of reversible transducers is straightforward.
Let us also remark that only the first transducer has to be reversible in order to obtain a polynomial complexity. 
However the reversible nature of the obtained transducer depends on the input transducers being both reversible.
\begin{theorem}\label{theorem::norm_composition}
Let $\tra_1$ and $\tra_2$ be two reversible two-way transducers with $n_1$ and $n_2$ states respectively, such that $\tra_1$ can be composed with $\tra_2$.
Then one can construct a reversible two-way transducer $\tra_3$ with $n_1\cdot n_2$ states realizing $\trans_{\tra_2}\circ\trans_{\tra_1}$.
\end{theorem}
\begin{proof}
Let $\tra_1=(A,B,Q,q_I,q_F,\Delta,\mu)$ and $\tra_2=(B,C,P,p_I,p_F,\Gamma,\nu)$.
We define $\tra_3=(A,C,Q\times P, (q_I,p_I),(q_F,p_F), \Theta,\xi)$.
The idea is that at each step, $\tra_3$ simulates a transition $\delta$ of $\tra_1$, plus
the behavior of $\tra_2$ over the production $\mu(\delta) \in B^*$ of this transition.
The partition of the set of states of $\tra_3$ depends on the combination of the signs of both components.
If $\tra_2$ is moving to the right, we use the determinism of $\tra_1$, we update the first component of the current state
according to the unique transition $\delta$ originating from it, and we simulate $\tra_2$ entering $\mu(\delta)$ from the left.
To do so, $\tra_3$ needs to have access to the same letter of the input tape as $\tra_1$.
Thus, we have $(Q^+\times P^+) \subseteq (Q\times P)^+$ and $(Q^-\times P^+) \subseteq (Q\times P)^-$.
If $\tra_2$ is moving to the left, then we use the co-determinism of $\tra_1$ to rewind the corresponding run, we update
the first component of the current state according to the unique transition $\delta$ arriving in it,
and we simulate $\tra_2$ entering $\mu(\delta)$ from the right.
To do so, $\tra_3$ needs to have access to the letter of the input tape opposite to $\tra_1$.
Thus, we have $(Q^-\times P^-) \subseteq (Q\times P)^+$ and $(Q^+\times P^-) \subseteq (Q\times P)^-$.
We now define the transition function $\Theta$ and the production function $\xi$.
Let $(q,a,q') \in \Delta$ be a transition of $\tra_1$ such that $\rho = (p,v,p')$ is an end-to-end run of $\tra_2$,
where $v$ denotes the word $\mu(q,a,q') \in B^*$.
\begin{itemize}
\item If $\rho$ is a left-to-right run of $\tra_2$, then $((q,p),a,(q',p'))$
	belongs to $\Theta$ and produces $\nu(p,v,p')$.
\item If $\rho$ is a left-to-left run of $\tra_2$, then $((q,p),a,(q,p'))$ belongs to $\Theta$ and produces $\nu(p,v,p')$.
\item If $\rho$ is a right-to-right run of $\tra_2$, then 
$((q',p),a,(q',p'))$ belongs to $\Theta$ and produces $\nu(p,v,p')$.
\item If $\rho$ is a right-to-left run of $\tra_2$, then
$((q',p),a,(q,p'))$ belongs to $\Theta$ and produces $\nu(p,v,p')$.
\end{itemize}
The behavior of the transducer $\tra_3$ is completely determined by the combined behaviors of transducers $\tra_1$ and $\tra_2$. 
When $\tra_3$ simulates a transition of $\tra_1$, it also simulates the corresponding end-to-end run of $\tra_2$ over the production of this transition. 
If the direction of both simulations is the same, then $\tra_3$ moves forward. Otherwise, it moves backward.
It ends when it has reached a final state of $\tra_1$ over the input, and a final state of $\tra_2$ over the sequence of partial productions of the run of $\tra_1$ over the input.
As a consequence, the transducer $\tra_3$ realizes the composition $\tra_2\circ\tra_1$. 
The determinism and co-determinism of $\tra_3$ is a direct consequence of the one of $\tra_1$ and $\tra_2$.
Indeed, a witness of non-determinism (resp. non co-determinism) of $\tra_3$ can be traced back to a witness run of either $\tra_1$ and $\tra_2$ that is not deterministic (resp.co-deterministic).
\end{proof}

\subsection{One-way transducers}
In the next subsections, we give some procedures to construct a reversible transducer from either a one-way or two-way transducer.
The main ingredient of the proofs is the technical construction from Lemma~\ref{lemma::tam1FT_to_norm2FT} (presented in Section~\ref{Section:proof}) which constructs a reversible transducer from a \emph{weakly branching} co-deterministic one-way transducer.
The proofs of this section share the same structure: in order to build
a reversible transducer that defines a function $\mathcal{F}$, we express $\mathcal{F}$
as a composition of transductions definable by reversible transducers,
and we conclude by using Theorem~\ref{theorem::norm_composition}.
The detailed constructions are presented in the appendix, for the sake of completeness.
Building on Lemma~\ref{lemma::tam1FT_to_norm2FT}, we show that co-deterministic one-way transducers can be expressed as the composition of weakly branching co-deterministic ones.
\begin{restatable}[]{theorem}{thmtwo}\label{theorem::cd1FT_to_norm2FT}
Given a co-deterministic \oFT{} with $n$ states, one can effectively construct an equivalent reversible \tFT{} with $4n^2$ states.
\end{restatable}

\begin{proof}
Let $\tra$ be a co-deterministic \oFT{} with $n$ states.
The function $\trans_{\tra}$ can be expressed
as the composition $\trans_{\tra'} \circ \trans_{\mult}$, where $\mult$ and $\tra'$
are defined as follows.
\begin{itemize}
\item
Transducer $\mult$ is a reversible \oFT{} with a single state that multiplies all the letters of the input word by $n$ while marking them with a state of $\tra$;
\item
Transducer $\tra'$ is a weakly branching and co-deterministic one-way transducer
that has the same set of states as $\tra$.
On input $\trans_{\mult}(u)$, $\tra'$ mimics the behavior of $\tra$ on $u$,
while using the fact that the input word is larger to desynchronize the non-deterministic branchings
that were occurring simultaneously in $\tra$. Intuitively, a transition of $\tra$ can only be taken by $\tra'$ at the copy of the letter corresponding to the target state of the transition.
\end{itemize}
By Lemma~\ref{lemma::tam1FT_to_norm2FT}, $\tra'$ can be made into a reversible \tFT{} ${\tra''}$ with $4n^2$ states.
Therefore, since both $\tra''$ and $\mult$ are reversible, we can conclude using Theorem~\ref{theorem::norm_composition}, finally obtaining a reversible \tFT{} with $4n^2$ states
equivalent to $\tra$.
\end{proof}

Using composition again, the statement can be extended to deterministic one-way transducers.
\begin{theorem}\label{theorem::1FT_to_norm2FT}
Given a deterministic \oFT{} with $n$ states, one can effectively construct an equivalent reversible \tFT{} with $36n^2$ states.
\end{theorem}
\begin{proof}
Let $\tra$ be a deterministic \oFT{} with $n$ states.
Then $\widebar{\tra}$, the transducer obtained by reversing all transitions of $\tra$, is co-deterministic.
The function $\trans_{\tra}$ can be expressed as the composition
$\trans_{M_{\alpo}} \circ \trans_{\widebar{\tra}} \circ \trans_{M_{\alp}}$, where
$M_{\alp}$ and $M_{\alpo}$ realize the mirror functions over the input and output alphabet of $\tra$ respectively. Both of them are realized by a $3$ states reversible transducer.
Then by Theorem~\ref{theorem::cd1FT_to_norm2FT}, we can construct $\widebar{\tra}'$ which has $4n^2$ states, is reversible and realizes the same function as $\widebar{\tra}$.
By Theorem~\ref{theorem::norm_composition}, we can compose the three transducers, finally obtaining a reversible transducer equivalent to $\tra$ with $9\cdot 4n^2$ states.
\end{proof}
%
%
%
\subsection{Two-way transducers}
We now prove our main result, which states that any two-way transducer can be uniformized by a reversible 
two-way transducer. Let us recall that uniformization by a deterministic transducer was done in~\cite{DS13}.
 We use similar ideas for the uniformization. The key difference is that we rely on the construction of Section~\ref{Section:proof} while in~\cite{DS13}, the main construction is the tree-trimming construction of Hopcroft-Ullman from~\cite{HU67}.
\begin{theorem}\label{theorem::2FT_to_norm2FT}
Given a \tFT{} $\tra$ with $n$ states, one can effectively construct a reversible \tFT{} $\tra'$ whose
number of states is exponential in $n$, and verifying $\lang_{\aut_{\tra'}} = \lang_{\aut_{\tra}}$ and
$\trans_{\tra'} \subseteq \trans_{\tra}$.
\end{theorem}

\begin{proof}
Let $\tra = (\alp,\alpo, Q, q_I, q_F, \Delta, \mu)$ be a \tFT{} with $n$ states.
We define a function uniformizing $\trans_{\tra}$ as the composition
$\trans_{\tra'} \circ \trans_{\unif} \circ \trans_{\odet_{r}}$,
where $\odet_{r}$, $\unif$ and $\tra$ are defined as follows.
\begin{itemize}
\item
The right-oracle $\odet_{r}$ is a co-deterministic one-way transducer with $2^{n^2 + n}$ states
that enriches each letter of the input word $u \in \vdashv{\alp}^*$ with information concerning
the behavior of $\tra$ on the corresponding suffix, represented by the set of pairs that admit a left-to-left
run, and the set of states from which $\tra$ can reach the final state.
\item
The uniformizer $\unif$ is a deterministic one-way transducer with $n!$ states.
On input $u' = \trans_{\odet_r}(u)$, $\unif$ uses the information provided by $\odet_{r}$
to pick a run $\rho_u$ of $\tra$ on input $u$, and enriches each
letter $a_i$ of the input word with the sequence of transitions occurring in the run $\rho_u$ that
correspond to the letter $a_i$.
\item
Finally, the reversible transducer $\tra'$ has the same set of states as $\tra$, and follows the instructions
left by $\unif$ to solve the non-determinism and the non-co-determinism.
\end{itemize}
As a consequence of Theorem \ref{theorem::cd1FT_to_norm2FT} and Theorem \ref{theorem::1FT_to_norm2FT},
there exist two reversible \tFT{} ${\odet_{r}}'$ and ${\unif}'$ whose number of states are exponential in $n$,
and that verify ${\trans_{\odet_{r}'}} = \trans_{\odet_{r}}$ and $\trans_{{\unif}'} = \trans_{\unif}$.
Therefore, since ${\odet_{r}}'$, ${\unif}'$ and $\tra'$ are reversible, by Theorem
\ref{theorem::norm_composition} there exists a reversible transducer ${\tra}''$
whose number of states is exponential in $n$, and that satisfies
$\trans_{{\tra}''} = \trans_{\tra'} \circ \trans_{{\unif}'} \circ \trans_{{\odet_{r}}'} = \trans_{\tra}$.
\end{proof}
The following result is a direct corollary of Theorem~\ref{theorem::2FT_to_norm2FT}, applied to deterministic two-way transducers.
\begin{corollary}
Reversible two-way transducers are as expressive as deterministic two-way transducers.
\end{corollary}

\section{The tree-outline construction}\label{Section:proof}
In this section lies the heart of our result.
We show that any \emph{weakly branching} and co-deterministic transducer can be made reversible. These hypotheses allows us to simplify our proof, and still obtain a more general result, as a corollary.

\begin{lemma}
\label{lemma::tam1FT_to_norm2FT}
Let $\tra$ be a co-deterministic and weakly branching \oFT{} with $m$ states.
Then one can effectively construct a reversible \tFT{} $\tra'$ with $4{m}^2$ states that is equivalent to $\tra$.	
\end{lemma}
\begin{proof} The construction of this proof is illustrated on an example in Figure~\ref{fig:proof1FTt2FT}.
	Let $\tra =  (\alp, Q, q_I, q_F, \Delta,\mu)$ be a co-deterministic \oFT, and let $\prec$ be a total order over $Q$.
	Take as an example the co-deterministic \oFT $\tra$ presented in figure \ref{fig:1}.
	
	Let $\tra' =  (\alp, \state, f_I, f_F, \Delta',\mu')$ be a \tFT{} defined as follows:
	
	On input $u \in \lang_{\tra}$, $\tra'$ explores depth first  the run-tree $T_u$ composed of the initial runs of $\tra$ on the word $\vdash u \dashv$  (illustrated in Figure~\ref{fig:runTree}).
	More precisely it explores the ``sheath'' of the run-tree (see Figure~\ref{fig:sheath} for a graphical representation).
	To do this, the states of $\tra'$ are composed of two states of $\tra$ with a marker. The first state represents the upper part of the sheath, while the second state represents the lower part. Moreover the marker is used to denote whether we are above the branch ($\ud q$) or below the branch ($\ov q$).
	
	Initially we start with the state $\displaystyle\pair{\ud {q_I}}{\ov{q_I}}$ and go forward according to the transitions of $\tra$.
	While moving forward whenever a branching state $q$ is reached, if the state is marked $\ud q$ it moves to the maximal successor of $q$ (in order to stay above the branch) and symmetrically if the state is marked $\ov q$ it moves to the minimal successor of $q$ (in order to stay below the branch).
	Whenever one of the branch reaches a dead end we continue the sheath exploration by switching the marker (\emph{i.e.} changing from above the branch to below or vice-versa) and start moving backward accordingly to the transitions of $\tra$.
	While moving backward, if the successor of a branching state $q$ is reached, while we were inside the fork, \emph{e.g.} in state $\ov{q_{\max}}$ (where $q_{\max}$ is the maximal successor of $q$), we continue the exploration of the sheath by going in the state $\ud{q_{\min}}$ and we start moving forward again.
	 Whenever the upper and lower explorations of the sheath coincide, \emph{i.e.} in states of the form $\displaystyle\pair{\ud {q}}{\ov{q}}$ (represented in red in Figure~\ref{fig:runOfTp}), it means we are on a prefix of the accepting run, we can thus produce the corresponding output.

	 \begin{figure}[htbp]
	\begin{subfigure}{0.45\textwidth} 	 
	\begin{tikzpicture}[scale=0.7]
	\node[circle,draw,inner sep=4] (n0) at (0,0) {$2$} ;
	\node[circle,draw,inner sep=4] (n1) at (2,0) {$1$} ;
	\node[circle,draw,inner sep=4] (n2) at (4,0) {$0$};
	\draw[>=stealth]  (-1,0) ->node[midway,above] {$\vdash$} (n0);
	\draw[>=stealth]  (2,-1) ->node[midway,right] {$\vdash$} (n1);
	\draw[>=stealth]  (n2) ->node[midway,above] {$\dashv$} (5,0);
	\draw (n0) edge  node[midway,above] {$b$} (n1);
	\draw (n1) edge  node[midway,above] {$a,b$} (n2);
	\draw (n0) edge[loop above] node[midway, above]{$a$} (n0);
	\draw (n1) edge[loop above] node[midway, above]{$a$} (n1);
	\end{tikzpicture}
	\caption{A co-deterministic transducer $\tra$\label{fig:1}}
\end{subfigure}\hfill
	 	\begin{subfigure}{0.45\textwidth} 
	 		\resizebox{\textwidth}{!}{
	 			\begin{tikzpicture}
	 			\node (q1) {$q_I$};
	 			\node (q2) [right of=q1,yshift=1cm] {$2$};
	 			\node (q3) [right of=q1,yshift=-0cm] {$1$};
	 			\node (q4) [right of=q3,yshift=-1cm] {$0$};
	 			\node (q5) [right of=q3,yshift=0cm] {$1$};
	 			\node (q8) [right of=q5] {$0$};
	 			\node (qf) [right of=q8] {$q_F$};
	 			\node (q6) [right of=q2] {$2$};
	 			\node (q7) [right of=q6] {$1$};
	 			\path
	 			(q1) edge[->] (q2)
	 			(q1) edge[->,color = red] (q3)
	 			(q2) edge[->] (q6)
	 			(q6) edge[->] (q7)
	 			(q3) edge[->] (q4)
	 			(q3) edge[->,color = red] (q5)
	 			(q5) edge[->,color = red] (q8)
	 			(q8) edge[->,color = red] (qf)	
	 			;
	 			\end{tikzpicture}
	 		}
	 		\caption{The run-tree of $\tra$ on $\vdash ab \dashv$\label{fig:runTree}} 
	 	\end{subfigure}
 	
 		\begin{subfigure}{0.45\textwidth} 
 			
 			\resizebox{\textwidth}{!}{
 		\begin{tikzpicture}
 		\node (q1) {$q_I$};
 		\node (q2) [right of=q1,yshift=2cm] {$2$};
 		\node (q3) [right of=q1] {$1$};
 		\node (q4) [right of=q3,yshift=-2cm] {$0$};
 		\node (q5) [right of=q3] {$1$};
 		\node (q8) [right of=q5] {$0$};
 		\node (qf) [right of=q8] {$q_F$};
 		\node (q6) [right of=q2] {$2$};
 		\node (q7) [right of=q6] {$1$};
 		\path
 		(q1) edge[->] (q2)
 		(q1) edge[->,color = red] (q3)
 		(q2) edge[->] (q6)
 		(q6) edge[->] (q7)
 		(q3) edge[->] (q4)
 		(q3) edge[->,color = red] (q5)
 		(q5) edge[->,color = red] (q8)
 		(q8) edge[->,color = red] (qf)	
 		;
 		
 		\coordinate (u1) at ($(q1)+(0,0.3)$) ;
 		\coordinate (u2) at ($(q2)+(0,0.3)$) ;
 		\coordinate (u3) at ($(q6)+(0,0.3)$) ;
 		\coordinate (u4) at ($(q6)+(0,0.6)$) ;
 		\coordinate (u5) at ($(q6)+(0,0.9)$) ;
 		\coordinate (u6) at ($(q7)+(0.2,0.3)$) ;
 		\coordinate (u7) at ($(q7)+(0.2,-0.3)$) ;
 		\coordinate (u8) at ($(q6)+(0,-0.3)$) ;
 		\coordinate (u9) at ($(u8)+(0,-0.3)$) ;
 		\coordinate (u10) at ($(u9)+(0,-0.3)$) ;
 		\coordinate (u11) at ($(q2)+(0,-0.3)$) ;
 		\coordinate (u12) at ($(q3)+(0,0.3)$) ;
 		\coordinate (u13) at ($(q5)+(0,0.3)$) ;
 		\coordinate (u14) at ($(u13)+(0,0.3)$) ;
 		\coordinate (u15) at ($(u14)+(0,0.3)$) ;
 		\coordinate (u16) at ($(q8)+(0,0.3)$) ;
 		\coordinate (u17) at ($(qf)+(0,0.3)$) ;
 		\path
 		(u1) edge[-] (u2)
 		(u2) edge[-] (u3)
 		(u3) edge[-,bend right] (u4)
 		(u4) edge[-,bend left] (u5)
 		(u5) edge[-] (u6)
 		(u6) edge[-,bend left] (u7)
 		(u7) edge[-] (u8)
 		(u8) edge[-,bend right] (u9)
 		(u9) edge[-,bend left] (u10)
 		(u10) edge[-] (u11)
 		(u11) edge[-,bend right] (u12)
 		(u12) edge[-] (u13)
 		(u13) edge[-,bend right] (u14)
 		(u14) edge[-,bend left] (u15)
 		
 		(u15) edge[-] (u16)
 		(u16) edge[-] (u17)
 		;
 		
 		\coordinate (l1) at ($(q1)-(0,0.3)$) ;
 		\coordinate (l2) at ($(q3)-(0,0.3)$) ;
 		\coordinate (l3) at ($(q4)-(-0.2,0.4)$) ;
 		\coordinate (l4) at ($(q4)+(0.2,0.4)$) ;
 		\coordinate (ll4) at ($(q3)+(0.3,0)$) ;
 		\coordinate (l5) at ($(q5)-(0,0.3)$) ;
 		\coordinate (l6) at ($(q8)-(0,0.3)$) ;
 		\coordinate (l7) at ($(q8)-(0,0.6)$) ;
 		\coordinate (l8) at ($(q5)-(0,0.6)$) ;
 		\coordinate (l9) at ($(l4)+(0.3,0)$) ;
 		\coordinate (l10) at ($(l3)+(0,-0.3)$) ;
 		\coordinate (l11) at ($(l2)+(0,-0.3)$) ;
 		\coordinate (l12) at ($(l11)+(0,-0.3)$) ;
 		\coordinate (l13) at ($(l10)+(0,-0.3)$) ;
 		\coordinate (l14) at ($(l9)+(0.3,0)$) ;
 		\coordinate (l15) at ($(l8)-(0,0.3)$) ;
 		\coordinate (l16) at ($(l7)-(0,0.3)$) ;
 		\coordinate (l17) at ($(qf)-(0,0.3)$) ;
 		\path
 		(l1) edge[-] (l2)
 		(l2) edge[-] (l3)
 		(l3) edge[-,bend right] (l4)
 		(l4) edge[-,bend left] (l5)
 		(l5) edge[-] (l6)
 		(l6) edge[-,bend left] (l7)
 		(l7) edge[-] (l8)
 		(l8) edge[-,bend right] (l9)
 		(l9) edge[-,bend left] (l10)
 		(l10) edge[-] (l11)
 		(l11) edge[-,bend right] (l12)
 		(l12) edge[-] (l13)
 		(l13) edge[-,bend right] (l14)
 		(l14) edge[-,bend left] (l15)
 		(l15) edge[-] (l16)
 		(l16) edge[-] (l17)
 		;
 		\end{tikzpicture}
 	}
 		\caption{Graphical representation of the run of $\tra'$\label{fig:sheath}} 
 		\end{subfigure}\hfill
 	\begin{subfigure}{0.45\textwidth}
	
	\begin{tikzpicture}[]
\node at (0.7,0.8) {$\vdash$};
\node at (2.1,0.8) {$a$};
\node at (3.5,0.8) {$b$};
\node at (4.9,0.8) {$\dashv$};

\providecommand\wA{}
\providecommand\hA{}
\providecommand\wB{}
\providecommand\hB{}

\renewcommand{\wA}{0.7cm}
\renewcommand{\hA}{0.6cm}
\renewcommand{\wB}{0.3cm}
\renewcommand{\hB}{0.6cm}

\node[draw= red,circle,inner sep=3] (n1) at (0.3,0) {\scriptsize $I$};

\node[draw, minimum width=\wA, minimum height=\hA] (n2) at (1.4,0) {};

\node[draw, minimum width=\wA, minimum height=\hA] (n3) at (2.8,0) {};

\node[draw, minimum width=\wA, minimum height=\hA](n4) at (2.8,-0.8) {};

\node[draw, minimum width=\wA, minimum height=\hA] (n5) at (2.8,-1.6) {};

\node[draw, minimum width=\wA, minimum height=\hA] (n6) at (4.2,-1.6) {};

\node[draw, minimum width=\wA, minimum height=\hA] (n7) at (4.2,-2.4) {};

\node[draw, minimum width=\wA, minimum height=\hA] (n8) at (2.8,-2.4) {};

\node[draw, minimum width=\wA, minimum height=\hA] (n9) at (2.8,-3.2) {};

\node[draw, minimum width=\wA, minimum height=\hA] (n10) at (2.8,-4) {};

\node[draw, minimum width=\wA, minimum height=\hA] (n11) at (1.4,-4) {};

\node[draw, color = red, minimum width=\wA, minimum height=\hA] (n12) at (1.4,-4.8) {};

\node[draw, minimum width=\wA, minimum height=\hA] (n13) at (2.8,-4.8) {};

\node[draw, minimum width=\wA, minimum height=\hA] (n14) at (2.8,-5.6) {};

\node[draw, color = red, minimum width=\wA, minimum height=\hA] (n15) at (2.8,-6.4) {};

\node[draw, color = red, minimum width=\wA, minimum height=\hA] (n16) at (4.2,-6.4) {};

\node[draw= red,circle,inner sep=3] (n17) at (5.3,-6.4) {\scriptsize $F$};


\draw (n1) edge (n2);
\draw (n2) edge (n3);
\draw (n3) edge[out=0, in=0] (n4);
\draw (n4) edge[out=180, in=180] (n5);
\draw (n5) edge (n6);
\draw (n6) edge[out=0, in=0] (n7);
\draw (n7) edge(n8);
\draw (n8) edge[out=180, in=180] (n9);
\draw (n9) edge[out=0, in=0] (n10);
\draw (n10) edge (n11);
\draw (n11) edge[out=180, in=180] (n12);
\draw (n12) edge (n13);
\draw (n13) edge[out=0, in=0] (n14);
\draw (n14) edge[out=180, in=180] (n15);
\draw (n15) edge (n16);
\draw (n16) edge (n17);


\node at (1.4,+0.) {\scriptsize{$\staf{\live{2}}{\dead{1}}$}};

\node at (2.8,+0) {\scriptsize{$\staf{\live{2}}{\dead{0}}$}};

\node at (2.8,+0-0.8) {\scriptsize{$\staf{\live{2}}{\live{0}}$}};

\node at (2.8,-1.6+0) {\scriptsize{$\staf{\live{2}}{\dead{1}}$}};

\node at (4.2,-1.6+0) {\scriptsize{$\staf{\live{1}}{\dead{0}}$}};

\node at (4.2,-2.4+0) {\scriptsize{$\staf{\dead{1}}{\dead{0}}$}};

\node at (2.8,-2.4+0) {\scriptsize{$\staf{\dead{2}}{\dead{1}}$}};

\node at (2.8,+0-3.2) {\scriptsize{$\staf{\dead{2}}{\live{0}}$}};

\node at (2.8,+0-4) {\scriptsize{$\staf{\dead{2}}{\dead{0}}$}};

\node at (1.4,-4+0) {\scriptsize{$\staf{\dead{2}}{\dead{1}}$}};

\node at (1.4,-4.8 + 0) {\scriptsize{$\staf{\live{1}}{\dead{1}}$}};

\node at (2.8,-4.8+0) {\scriptsize{$\staf{\live{1}}{\dead{0}}$}};

\node at (2.8,-5.6+0) {\scriptsize{$\staf{\live{1}}{\live{0}}$}};

\node at (2.8,-6.4+ 0) {\scriptsize{$\staf{\live{1}}{\dead{1}}$}};

\node at (4.2,-6.4 + 0) {\scriptsize{$\staf{\live{0}}{\dead{0}}$}};


\end{tikzpicture}
	\caption{The run of $\tra'$\label{fig:runOfTp}}
\end{subfigure}
	 	\caption{Illustrations of the proof concepts\label{fig:proof1FTt2FT}}
	 \end{figure}
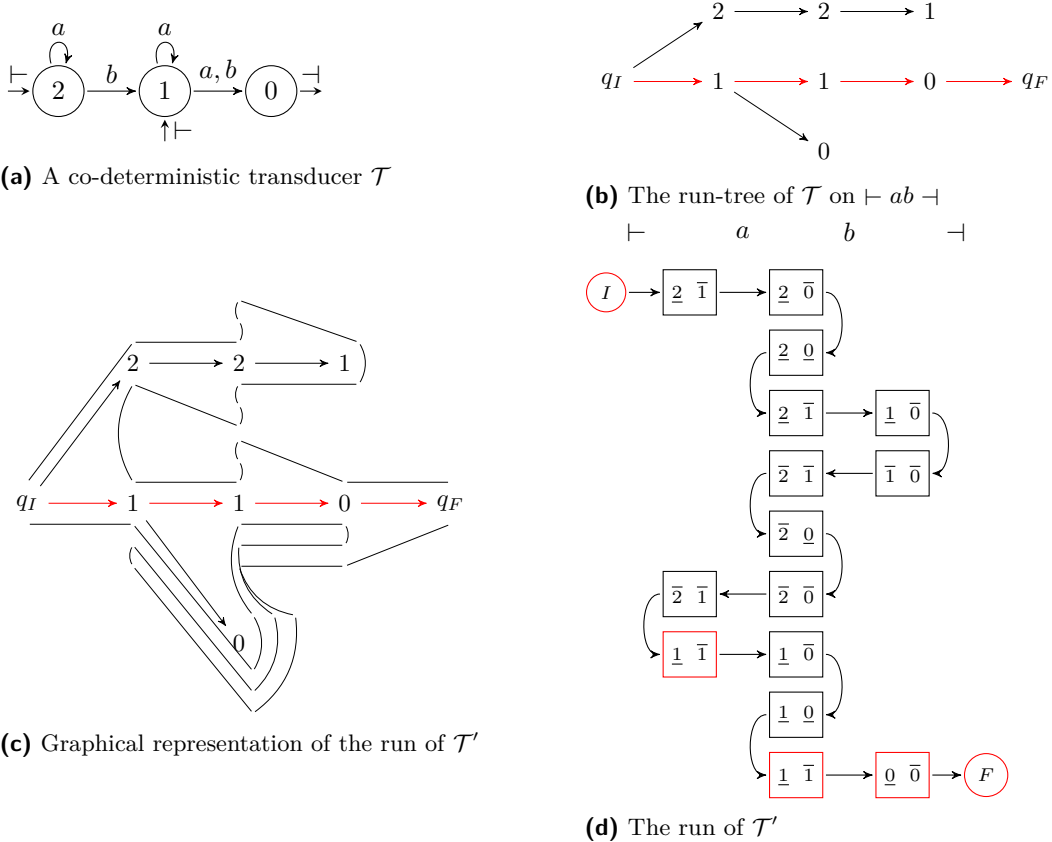
	
	Formally $\tra' =  (\alp, \state, f_I, f_F, \Delta',\mu')$ is defined as follows:
	\begin{itemize}
		\item $\state=\state^+\cup\state^-$ where 
		$\state^+=\ud{Q}\times \ov{Q}\cup \ov{Q}\times \ud{Q}$ and $\state^-=(\ov{Q}\times \ov{Q}\cup \ud{Q}\times\ud{Q})\setminus\{(\ov p,\ov p),(\ud p,\ud p)\mid p\in Q\}$
		\item $\displaystyle f_I=\pair{\ud{q_I}}{\ov{q_I}}$
		\item $\displaystyle f_F=\pair{\ud{q_F}}{\ov{q_F}}$
		\item We define the transition relation $\Delta'$ by differentiating several types of behavior, 
		depending on whether we are going \textbf{f}orward, or \textbf{b}ackward, whether the \textbf{u}pper component or the \textbf{l}ower component is involved, and whether it is \textbf{a}bove or belo\textbf{w} its branch.
		Let $p$ and $q$ be two states in $Q$, and $a\in\alp$ be a letter.
		
		If $p$ has no $a$-successor, then:
			\begin{description}
				\item[(fua)] $\trs{\ud{p}}{\ov{q}}{\ov{p}}{\ov{q}}\in\Delta'$, and
				\item[(fuw)] $\trs{\ov{p}}{\ud{q}}{\ud{p}}{\ud{q}}\in\Delta'$.
			\end{description}
			If $p$ has an $a$-successor, but not $q$, then:
			\begin{description}	
				\item[(flw)] $\trs{\ud p}{\ov q}{\ud p}{\ud{q}}\in\Delta'$, and
				\item[(fla)] $\trs{\ov{p}}{\ud{q}}{\ov{p}}{\ov q}\in\Delta'$.	
			\end{description}

		Otherwise, $p$ and $q$ admit an $a$-successor. We denote $p_{\max}$ (resp. $p_{\min}$) the maximal (resp. minimal) $a$-successor of $p$ (resp. $q$) with respect to $\prec$. Then:\\
		\begin{minipage}{0.45\textwidth}
			If $p_{\min}\neq p_{\max}$, then:
			\begin{description} 
				\item[(buw)] $\trs{\ov{p_{\max}}}{\ov q}{\ud{p_{\min}}}{\ov q}\in\Delta'$, and
				\item[(bua)] $\trs{\ud{p_{\min}}}{\ud{q}}{\ov{p_{\max}}}{\ud{q}}\in\Delta'$.
			\end{description}
			If $q_{\min}\neq q_{\max}$, then:
			\begin{description}
				\item[(bla)] $\trs{\ud p}{\ud{q_{\min}}}{\ud p}{\ov{q_{\max}}}\in\Delta'$
				\item[(blw)] $\trs{\ov{p}}{\ov{q_{\max}}}{\ov{p}}{\ud{q_{\min}}}\in\Delta'$
			\end{description}
		\end{minipage}\hfill
		\begin{minipage}{0.45\textwidth}
			\begin{description}
				\item[(fualw)] $\trs{\ud p}{\ov q}{\ud{p_{\max}}}{\ov{q_{\min}}}\in\Delta'$,
				\item[(fuwla)] $\trs{\ov{p}}{\ud{q}}{\ov{p_{\min}}}{\ud{q_{\max}}}\in\Delta'$,
				\item[(bulw)] $\trs{\ov{p_{\min}}}{\ov{q_{\min}}}{\ov{p}}{\ov q}\in\Delta'$, and
				\item[(bula)]  $\trs{\ud{p_{\max}}}{\ud{q_{\max}}}{\ud p}{\ud{q}}\in\Delta'$.
			\end{description}
		\end{minipage}
		\item We define $\mu'$ as the function such that for every $(p,a,q)\in \Delta$:
		\begin{itemize}
			\item if $q=p_{\min}=p_{\max}$ then $\mu'\trs{\ud p}{\ov p}{\ud q}{\ov q}=\mu(p,a,q)$
			\item if $q=p_{\min}\neq p_{\max}$ then $\mu'\trs{\ov{p_{\max}}}{\ov q}{\ud q}{\ov q}=\mu(p,a,q)$
			\item if $q=p_{\max}\neq p_{\min}$ then $\mu'\trs{\ud q}{\ud{p_{\min}}}{\ud q}{\ov q}=\mu(p,a,q)$
		\end{itemize}  
	and $\mu'(t)=\epsilon$ for every $t\in\Delta'$ which is not of one of theses forms.
	\end{itemize}
One can see, by a case study that $\tra'$ is deterministic. Indeed, the fact that $\tra$ is weakly branching implies that the rules \buw{} and \bua{} are mutually exclusive with the rules \bla{} and \blw{}. Moreover these four rules are mutually exclusive with the rules \bulw{} and \bula{} by construction. And since $\tra$ is co-deterministic, the predecessor is unique. Finally, the rules \fua{}, \fuw{}, \flw{}, \fla{}, \fualw{}, and \fuwla{} are mutually exclusive by construction, since the conditions on the number of $a$-successors are incompatible.

A similar case study gives that $\tra'$ is co-deterministic. Hence $\tra'$ is reversible. 

A detailed proof of the equivalence between $\tra$ and $\tra'$ can be found in the appendix, and we give a quick intuition of the proof.
It relies on two main arguments. 
The first one is that at any point if the transducer $\tra'$ follows two differents runs, then it will come back to the same position, where the state that leads to the shortest run has been switched.
Following this, we then prove that upon any branching, $\tra'$ comes back to the same position but since the shortest run has been switched, it is able to solve the non-determinism, take the transition of the accepting run and produce the correct output.
\end{proof}

\section{Streaming string transducers}\label{Section:sst}
Streaming string transducers, which were introduced in~\cite{AC10}, are one-way deterministic automata with additional write-only registers.
Partial outputs are stored in the registers via register updates, and at the end of a run an output is produced using these registers.
Thus a \SST{} realizes a function over words, and it is known that they are as expressive as \tFT{}~\cite{AC10}.
Direct transformations from \SST{} to \tFT{} were already considered in~\cite{AFT12,DJR16}.
However, these constructions were exponential in the number of states (and linear in the number of registers). 
Using Theorem~\ref{theorem::1FT_to_norm2FT}, we are able to get a construction which is quadratic in the number of states (and also linear in the number of registers).
Before explaining the construction, let us formally define the \SST{}.
\myparagraph{Substitutions}
Given a finite alphabet $\alp$ and a finite set $\var$ of variables.
Let $\sub{\var}{\alp}$ denote the set of functions
$\sigma : \var \rightarrow (\var \cup \alp)^*$.
The elements of $\sub{\var}{\alp}$ are called \emph{substitutions}.
Any substitution $\sigma$ can be extended to range over both variables and letters of the output alphabet $\hat{\sigma} : (\var \cup \alp)^* \rightarrow (\var \cup \alp)^*$ by setting $\hat{\sigma}(a) = a$ for every $a \in \alp^*$ and $\hat{\sigma}(uv)=\hat{\sigma}(u)\hat{\sigma}(v)$ for $u,v,\in (\var\cup\alp)^*$.
This allows us to easily compose substitutions from $\sub{\var}{\alp}$ by defining $\sigma_2 \circ \sigma_1$ as the usual function composition $\hat{\sigma_2} \circ \sigma_1$.
We denote by $\idsub{\var}$ the identity element of $\sub{\var}{\alp}$, which maps every variable to itself,
and by $\empsub$ the substitution mapping every variable to $\epsilon$.
Given $n \in \mathbb{N}$, a substitution $\sigma$ is called \emph{$n$-bounded} if for every $X \in \var$, each variable $Y \in \var$ appears at most once in $\sigma(X)$.
A substitution $\sigma$ is called \emph{copyless} if it is $1$-bounded, and for every $Y \in \var$ there exists at most one $X \in \var$ such that $Y$ appears in $\sigma(X)$.

\myparagraph{Streaming string transducers}
A \emph{streaming string transducer} (\SST) is a tuple $\sst = (\alp,\alpo, Q, q_I, q_F,\Delta, \var,\varO, \tau)$,
where $\alpo$ is the output alphabet, $\aut_{\sst} = (\alp, Q, q_I, q_F, \Delta)$ is a one-way deterministic
automaton, called the underlying automaton of $\sst$;
$\var$ is a finite set of variables; 
$\varO \in \var$ is the final variable; 
$\tau : \Delta \rightarrow \sub{\var}{\alpo}$ is the output function.
A run of $\sst$ is a run of its underlying automaton, and the language $\lang_{\sst}$ recognized by $\sst$
is the language $\lang_{\aut_{\sst}} \in \alp^*$ recognized by its underlying automaton.
Given a run $\rho$ of $\sst$
on $u$, we set $\tau(\rho) \in \sub{\var}{\alpo}$ as the composition of the images by $\tau$ of the transitions of $\sst$ occuring
along $\rho$.
The transduction $\trans_{\sst} \subseteq \alp^* \times \alpo^*$ defined by $\sst$ is the function mapping any word $u$ of
$\lang_{\aut_{\sst}}$ to $(\sigma_{\epsilon} \circ \tau(\rho))(\varO)$, where $\rho$ is the single accepting run of $\aut_{\sst}$
on $\vdash u \dashv$.
The \SST $\sst$ is called $n$-bounded, respectively copyless, if for every run $\rho$ of $\sst$ the substitution $\tau(\rho)$ is
$n$-bounded, respectively copyless.

\begin{theorem}\label{theorem::SST_to_norm2FT}
Given a copyless \SST{} with $n$ states and $m$ variables, one can effectively construct an equivalent
reversible \tFT{} with $8m\cdot n^2$ states.
\end{theorem}

\begin{proof}
We write $\sst$ as the composition of a one-way deterministic transducer $\odet_1$ and a reversible one $\tra$.
The first transducer has the same underlying automaton as $\sst$, the difference being that it outputs the substitution of $\sst$ instead of applying it.
Then $\tra$ is a transducer that navigates the substitutions to produce the output word of $\sst$.
This can be done in a reversible fashion thanks to the property of copylessness of $\sst$.
Note that the transducer $\tra$ was already defined in~\cite{DJR16}, Section~4.
Formally, let $\sst = (\alp,\alpo, Q, q_I, q_F,\Delta, \var,\varO, \tau)$ be a copyless \SST{} with $n$ states and $m$ variables,
and let $S_{\sst} \subset \sub{\alp}{\var}$ be the range of $\tau$.
We express $\trans_{\sst}$ as the composition of $\trans_{\odet_1} : \lang_{\aut_{\sst}} \rightarrow S_{\sst}^*$
and $\trans_{\tra_2} : S_{\sst}^* \rightarrow \alpo^*$, defined as follows.
\begin{itemize}
\item
$\odet_1$ is a deterministic \oFT{} obtained by stripping $\sst$ of its \SST{} structure, i.e., 
$\odet_1 = (\alp,S_{\sst}, Q, q_I, q_F,\Delta, \tau)$.
It maps each word of $\lang_{\aut_{\sst}}$ to the corresponding sequence of substitutions.
\item
$\tra=(S_\sst,\alpo,P,init,fin,\Gamma,\nu)$ where $P^+=\var^{o}\uplus\{init,fin\}$, $P^-=\var^{i}$. States labeled by $i$ (resp. $o$) are \emph{in} (resp. \emph{out}) states and appear when we start (resp. finish) producing a variable.
We define $\Gamma$ and $\nu$ as follows:
	\begin{itemize}
	\item $(init,\sigma,init)\in\Gamma$;
	\item $(init,\dashv,O^i)\in\Gamma$;
	\item $(O^o,\dashv,fin)\in \Gamma$;
	\item $(X^i,\sigma,Y^i)\in\Gamma$ and $\nu((X^i,\sigma,Y^i))=v$ if $\sigma(X)=vY...$ with $v\in \alpo^*$;
	\item $(X^i,\sigma,X^o)\in\Gamma$ and $\nu((X^i,\sigma,X^o))=v$ if $\sigma(X)=v$;
	\item $(X^o,\sigma,Y^i)\in\Gamma$ and $\nu((X^o,\sigma,Y^i))=v$ if there exists a variable $Z$ where $\sigma(Z)=...XvY..$;
	\item $(X^o,\sigma,Y^o)\in\Gamma$ and $\nu((X^o,\sigma,Y^o))=v$ if $\sigma(Y)=...Xv$.
 	\end{itemize}
 	Due to copylessness, for any $\sigma$ and any variable $X$, there is at most one variable $Y$ such that $X$ appears in $\sigma(Y)$. Plus, as the variables are ordered by their appearance in $\sigma(Y)$, the transducer $\tra$ is reversible.
 	It starts by reaching the end of the word, then starts producing the variable $O$.
 	By following the substitution tree of $O$, it then produces exactly the image of the input by $\sst$.
\end{itemize}
By Theorem~\ref{theorem::1FT_to_norm2FT}, there exists a reversible \tFT{} $\odet_1'$ with $4n^2$
states satisfying $\trans_{\odet_1'} = \trans_{\odet_1}$.
Finally, since both $\odet_1'$ and $\tra$ are reversible, by Theorem~\ref{theorem::norm_composition}
there exists a reversible transducer $\tra'$
with $8m\cdot n^2$ states such that $\trans_{\tra'} = \trans_{\tra} \circ \trans_{\odet_1'} = \trans_{\sst}$.
\end{proof}

 \section{Conclusion}\label{Section:Concl}
We argue that reversible transducers can be seen as a canonical way to represent two-way transducers.
We believe that the polynomial complexity of composition of reversible transducers is a good tool for the verification of cascades of transformations of non-reactive systems.
While not restricting the expressive power, reversible transducers allow for the easiest manipulations, the best example being their polynomial composition.
Thanks to the tree-outline construction that we presented, one can uniformize a non-determinsitic two-way transducer into a reversible one with a single exponential blow-up.
While this improves the known construction that were used up to now, it is still open whether this blow-up can be avoided. 
In \cite{KuncO13} the authors extended the result of \cite{KondacsW97} and showed that deterministic two-way automata can be made reversible with a linear blow-up.
We conjecture that our approach can also be extended to the two-way case and that deterministic two-way transducers can be made reversible using only a polynomial number of states.

We showed that applying this construction allowed for a quadratic transformation from copyless streaming string transducers to reversible two-way transducers. 
The converse does not hold, since even on languages deterministic two-way automata are known to be exponentially more succint than deterministic one-way automata.
Beyond this, we argue that if one were to embed some recognition power into the variables of a \SST, it may be possible to have a polynomial transformation from reversible automata to copyless \SST.



\bibliography{biblio}


\newpage
\appendix
\noindent{\bf \Large Appendix}
\medskip 
\section{Proof of Theorem \ref{theorem::cd1FT_to_norm2FT}}
\thmtwo*
 
Let $\tra = (\alp, \alpo,Q,q_{I},q_{F},\Delta,\mu)$ be a co-deterministic
\oFT, with $n$ states.
We present here the detailed constructions of two one-way transducers $\mult$ and $\tra'$
such that $\mult$ is reversible, $\tra'$ is weakly branching and co-deterministic, and
$\trans_{\tra'} \circ \trans_{\mult} = \trans_{\tra}$.

Consider an ordering $q_1<\ldots<q_n$ of $Q$.
The transducer $\mult$ takes as input words $u$ from $\vdashv{\alp}^*$ and returns words from $((\vdashv{\alp}\times Q)\uplus\{r\})^*$. It basically includes $<$ into each letter of $u$ by copying each letter $|Q|$ times and adds in the end a letter $r$ for \emph{reset}.
Formally, let $\mult = (\vdashv{\alp}, (\vdashv{\alp}\times Q)\uplus\{r\},\{id\},id,id,\Delta_<,\mu_<)$, where 
$\Delta_<=\{(id,a,id)\mid a\in \vdashv{\alp}\}$ and 
$\mu_<$ maps any transition $(id,a,id)$ to $(a,q_1)\ldots(a,q_n)r$. Since $\mult$ only has one states, it is clearly reversible.

The transducer $\tra'$ is designed to take a word $\mult(u)$ as input and mimic the behavior of $\tra$ on $u$.
It spreads the simultaneous non-deterministic branchings of $\tra$ over the different copies of the letters, in order to be weakly branching.
The idea is that upon reading a letter $(a,q)$, the transducer can either do nothing or take a transition to the state $q$ if it existed in $\tra$.
In order to ensure that exactly one transition is taken while reading copies of the same letters, a counter is placed on the states that is incremented when a transition is taken, and reset uppon reading the letter $r$.
Formally, let now $\tra'=((\vdashv{\alp}\times Q)\uplus\{r\},\alpo,Q\times\{0,1\},q_I,q_F,\Delta',\mu')$ be defined on words of $((\vdashv{\alp}\times Q)\uplus\{r\})^*$ where $\Delta'$ contains the transitions:
\begin{itemize}
\item $((p,0),(a,q),(q,1))$ for all $a\in \vdashv{\alp}, p,q\in Q$ such that $(p,a,q)\in \Delta$,
\item $((p,0),(a,q),(p,0))$ for all $a\in \vdashv{\alp}, p,q\in Q$,
\item $((p,1),(a,q),(p,1))$ for all $a\in \vdashv{\alp}, p\neq q\in Q$,
\item $((p,1),r,(p,0))$ for all $p\in Q$.  
\end{itemize}

The function $\mu'$ matches the transitions $((p,0),(a,q),(q,1))$ to the corresponding production 
$\mu(p,a,q)$, and produces $\epsilon$ in the other cases.

We prove that the transducers $\mult$ and $\tra'$ satisfy the desired properties.
First, let us recall that a transducer is weakly branching if for each letter, there is at most one state that creates nondeterminism, and this nondeterminism is between two choices.
Notice that states of $\tra'$ labeled by $1$ are deterministic, and that sates labeled by $0$ 
can, upon reading a letter $(a,q)$, either stay in the same state or possibly go to state $(q,1)$. Then $\tra'$ nondeterminism appears between two states.
Consider a letter $(a,q)$. Nondeterminism on $(a,q)$ can only occur for transitions $(p,a,q)$ of $\tra$. 
Since $\tra$ is co-deterministic, for any $(a,q)$ there is at most one state $p$ such that $(p,a,q)$ is a transition of $\tra$. Hence, for a given letter $(a,q)$ there exists at most one state $p$ that can create nondeterminism.
Regarding co-determinism of $\tra'$, given a state $(p,0)$ its predecessor can only be $(p,0)$ upon reading a letter $(a,q)$ and $(p,1)$ upon reading $r$.
If we consider a state $(p,1)$ its predecessor upon reading a letter $(a,q)$ where $p\neq q$ has to be $(p,1)$, while upon reading $(a,p)$ it can only be a state $(q,0)$ such that $(q,a,p)$ is a transition of $\tra$.
Consequently, if $\tra$ is co-deterministic, then so is $\tra'$.

To conclude, we now prove that $\trans_{\tra}=\trans_{\tra'}\circ\trans_{\mult}$.
Consider a pair $(u,v)$ of $\trans_{\tra'}\circ\trans_{\mult}$.
As the image of $u$ by $\mult$ is $(u_1,q_1)\ldots(u_1,q_n)r\ldots(u_k,q_1)\ldots(u_k,q_n)r$, any accepting run $\rho$ of $\tra'$ on $(u_1,q_1)\ldots(u_1,q_n)r\ldots(u_k,q_1)\ldots(u_k,q_n)r$ can be traced back to a sequence of producing transitions $t_i=((p_i,0),(u_i,p_{i+1}),(p_{i+1},1))$ such that $v=\mu'(t_1)\ldots\mu'(t_k)$.
By construction, only one such transition can appear in a given word without $r$, and such transitions come from transitions $(p_i,u_i,p_{i+1})$ of $\tra$ and we have $v=\mu((p_1,u_i,u_2))\ldots\mu((p_k,u_k,q_F))$.
Thus $(u,v)$ is also a pair of $\trans_{\tra}$.
Conversely, a run of $\tra$ can be transformed into a run of $\tra'$. 
The transducer $\mult$ ensures that the input given to $\tra'$ can make this run.

\section{Proof of Theorem \ref{theorem::2FT_to_norm2FT}}

Let $\tra = (\alp, \alpo,Q,q_{I},q_{F},\Delta,\mu)$ be a
\tFT, and let $n = |Q|$.
We present here the detailed constructions of the transducers $\unif$,  $\odet_{r}$ and $\tra'$
used in the proof of Theorem \ref{theorem::2FT_to_norm2FT}.
We begin by introducing new definitions and notations.

\myparagraph{Ordering the runs of $\tra$}
Let $\prec$ be a total order on $Q$.
First, using $\prec$, we define the \emph{length-lexicographical} order $\preclex$ on the set $Q^*$ of finite sequences of states.
Formally, we have $p_0 \ldots p_{k} \preclex q_0 \ldots q_{k'}$ if either $k < k'$, or $k = k'$ and there exists an index $0 \leq i \leq k$
verifying $p_i \prec q_i$ and $p_j = q_j$ for every $0 \leq j < i$.
Second, using $\preclex$, for every $u \in \vdashv{\alp}^*$ we define a lexicographical order $\precsli$
on the accepting runs of $\tra$ on $u$.
For every run $\rho$ on $u$ and any prefix $u_1$ of $u$, consider the corresponding subsequence
$u_1.q_{i_1}.u_2,u_1.q_{i_2}.u_2, \ldots, u_1.q_{i_k}.u_2$ of configurations of $\rho$ such that the reading head is positioned right after $u_1$.
Then the \emph{slice} $\projrun{|u_1|}{\rho} \in Q^*$ is equal to the sequence $q_{i_1}q_{i_2}, \ldots q_{i_k} \in Q^*$.
We say that $\rho \precsli \rho'$ if there exists a prefix $v$ of $u$ verifying $\projrun{|v|}{\rho} \preclex \projrun{|v|}{\rho'}$, and
$\projrun{|v'|}{\rho} = \projrun{|v'|}{\rho'}$ for every prefix $v'$ of $v$.
We denote by $\rho_u$ the minimal accepting run on $u$ with respect to $\precsli$.
An accepting run $\rho$ of $\tra$ on $u$ is called \emph{irreducible} if no subsequence of configurations of $\rho$ is an accepting run.
Note that, by minimality, $\rho_u$ is irreducible.
Therefore, the same configuration is never repeated twice along $\rho_u$, and the length of the slices of $\rho_u$ is bounded by $n$.
This order on runs is used by the second transducer $\unif$ (the \emph{uniformizer}) which selects the minimal accepting run of $\tra$ given the information from the right oracle $\odet_r$.

\myparagraph{Construction of the right oracle}
For every word $w \in \vdashv{\alp}^*$, we represent the behavior of $\tra$ on $w$ starting from the left with a pair
$\lbeh{w} = (\llbeh{w},\predfin{w})  \in 2^{Q \times Q} \times 2^{Q}$.
Formally, $\llbeh{w} \subseteq Q \times Q$ denotes the left-to-left runs on $w$, i.e. the set of pairs $(p,q) \in Q^+ \times Q^-$
satisfying $(p,w,q) \in \Delta$, and $\predfin{w} \subseteq Q$ denotes the set of $w$-predecessors
of the final state $q_F$.
On input $u \in \lang_{\tra}$, the right oracle $\odet_{r} = (\alp, \alp_r,Q_{r},{I}_{r},{F}_{r},\Delta,\mu_r)$
enriches each letter of $u$ with the behavior of $\tra$ on the corresponding suffix.
It has the following components. 
\begin{itemize}
\item The output alphabet $\alp_r$ is equal to the product $\alp \times 2^{Q \times Q} \times 2^{Q}$;
\item the set of states $Q_{r}$ is composed of the left behaviors $\lbeh{w}$, for every $w \in \vdashv{\alp}^*$;
\item the initial state is $I_{r} = (\emptyset,\{q_I\})$;
\item the final state is $F_{r} = (\emptyset,\{q_F\})$;
\item the transition relation $\Delta_{r}$ contains the triples $(\lbeh{aw},a,\lbeh{w})$, for all $w \in \vdashv{\alp}^*$, $a \in \vdashv{\alp}$, and the triples $(I_r,\vdash,\lbeh{w})$, for all $w\in\vdashv{\alp}^*$ where there exists $q\in \predfin{w}$ such that $(q_I,\vdash,q)\in\Delta$;
\item the output function $\mu_{r} : \Delta \rightarrow \alpo^*$ maps $(\lbeh{aw},a,\lbeh{w}) \in \Delta_r$ to $(a,\lbeh{w}) \in \alp_r$.
\end{itemize}

In order to prove that the transition relation is computable and that $\odet_{r}$ is co-deterministic,
we expose the construction of $\lbeh{aw}$ from $a$ and $\lbeh{w}$. 
This comes from the fact that every run of $\tra$ on $aw$ can be expressed as 
the concatenation of runs on $w$ and transitions corresponding to the letter $a$.
Formally, let $ Cl(\llbeh{w})\subseteq Q^+ \times Q^-$ be the set of pairs $(p_0,p_k)$ such that there exists
$p_1, \ldots, p_{k-1} \in Q^+$, $q_1, \ldots, q_{k} \in Q^-$ verifying $(p_i,q_{i+1}) \in \llbeh{w}$
for $0\leq i<k$ and $(q_i,a,p_{i}) \in \Delta$ for every $1 \leq i < k$.
Then $(p,q) \in \llbeh{aw}$ if and only if either $(p,a,q) \in \Delta$, or there exists $(p',q') \in Cl(\llbeh{w})$ such that
$(p,a,p'),(q',a,q) \in \Delta$.
Moreover, $p \in \predfin{aw}$ if and only if either there exists $q\in\predfin{w}$ such that $(p,a,q)\in\Delta$ or there exists $(p',q') \in Cl(\llbeh{w})$, $q''\in Q^+$ such that $(p,a,p'),(q',a,q'') \in \Delta$ and $q'' \in \predfin{w}$.

\myparagraph{Construction of the uniformizer}
On input $\odet_{r}(u)$, for some $u \in \lang_{\tra}$, the transducer $\unif$ uses the information provided by $\odet_{r}$ to determine
the sequence of slices corresponding to the minimal accepting run $\rho_u$ of $\tra$ on $u$.
The set of states $Q_{sl}$ of $\unif$ is the set of sequences of states of $Q$ of size less than or equal to $n$.
We define 
$\unif = (\alp_r, \alp_{sl},Q_{sl},i,f,\Delta_{sl},\mu_{sl})$ as follows:
\begin{itemize}
\item The output alphabet $\alp_{sl}$ consists of $\Delta^{\leq n}$, the bounded sequences of transitions;
\item the set of states $Q_{sl}$ is composed of the sequences of  $Q^{*}$ whose length is bounded by $n$;
\item the transition relation $\Delta_{sl}$ contains the triples $(\projrun{|u|}{\rho_{uav}},(a,\lbeh{v}),\projrun{|ua|}{\rho_{uav}})$, for all $u,v \in \vdashv{\alp}^*$, $a \in \vdashv{\alp}$;
\item the output function $\mu_{sl} : \Delta \rightarrow \alpo^*$ maps $(\projrun{|u|}{\rho_{uav}},(a,\lbeh{v}),\projrun{|ua|}{\rho_{uav}})$ to the sequence $t_1\ldots t_k$ of transitions reading $a$ in $\rho_{uav}$.
\end{itemize}

We now prove that the transition relation is computable, and that $\unif$ is deterministic, by
constructing $\projrun{|ua|}{\rho_{uav}}$ from $a$, $\lbeh{v}$ and $\projrun{|u|}{\rho_{uav}}$.
Once again, we use the fact that every run of $\tra$ on $av$ can be expressed as the concatenation of runs on $v$
and transitions corresponding to the letter $a$.
The main difficulty is to locally identify the sequence that corresponds to the slice of the minimum run.
By definition of the minimal run, this amounts to always select the minimal slice that is compatible with the previous information, and is valid, i.e. can be extended to a whole run over the input.
Since we have access to the left-to-left behavior of the suffix, we know which slices are valid.
And thanks to the current state, we have access to the last slice, and so we know which slice can be composed with it.

Formally, given a slice $\pi=p_1\ldots p_k$, we denote by $\pi^+$ (resp. $\pi^-$) the subsequence of $\pi$ of states from $Q^+$ (resp. $Q^-$).
Let us construct the set of slices that are compatible and valid with $\pi$ upon reading some letter $(a,(R,F))$.
For every $p_i$ in $\pi^-$, $(p_i,p_{i+1})$ describes a left-to-left run of the prefix.
Now let us determine the behavior of the states $p_i$ of $\pi^+$ on the current letter.
As we aim to construct the minimal slice that is compatible with $\pi$, if $(p_i,a,p_{i+1})$ belongs to $\Delta$ then the minimal slice takes this transition.
Thus we can precisely identify which states $p_i$ crosses the letter $a$ and take a transition to a state of the slice $\pi'$ we are constructing.
Then amongst all slices, we can identify the ones that are compatible with $\pi$, i.e. the ones that have a state reachable from $p_i$, such that the next state that belongs to $Q^-$ has a transition to $p_{i+1}$.
Within this set of slices, for each slice $\pi'$ we can check if it is valid with respect to $(R,F)$: 
it suffices to verify that for all state $q_j$ of $\pi'^+$, $(q_j,q_{j+1})\in R$.
We also verify that the last state of $\pi'$ belongs to $F$.
Thus the set of compatible and valid slices is computable and finite, and we can chose the smallest one with respect to $\preceq_{sl}$.
Moreover, since we identified how the two slices are linked, we know exactly which transitions are taken across the letter $a$ and their relative order in the run, and we can output the sequence of transitions relative to $a$.

\myparagraph{Construction of the reversible transducer}
The last transducer $\tra'$ simply reads the slices and follows the run described by the production of $\unif$.
Its set of states corresponds to the one of $\tra$, and is used to situate the run in the current slice.
Formally, we set $\tra'=(\alp_{sl},B,Q,q_I,q_F,\Delta',\mu')$ where
the transition relation $\Delta'$ is the set of triplets $(p,(t_1\ldots t_k),p')$ such that 
there exists $t_i=(p,a,p')\in \Delta$
and we set $\mu'((p,(\pi,a,\pi'),p'))=\mu((p,a,p')$.

Since in a slice of the minimal run, states are not repeated, each state appear at most one time as a left component and right component of a transition of $t_1\ldots t_k$, we hence have determinism and co-determinism of $\tra'$.
Moreover, as for each input $u$ of $\lang_{\aut_{\tra}}$ we selected the minimal accepting run of $\tra$, we finally get that the composition $\tra'\circ\unif\circ\odet_r$ is a uniformization of $\tra$.

\myparagraph{Conclusion}
In the end, we have that  $\tra'\circ\unif\circ\odet_r$ is a uniformization of $\tra$, where $\tra'$ has $n$ states, $\unif$ has $n!$ states and $\odet_r$ has $2^{n^2+n}$ states.
Using Theorems~\ref{theorem::cd1FT_to_norm2FT}  and~\ref{theorem::1FT_to_norm2FT},
we can construct some reversible transducers $\odet_r'$ and $\unif'$ that respectively have $4\cdot 2^{2(n^2+n)}=2^{2(n^2+n+1)}$ and $36(n!)^2$ states.
Finally, by Theorem~\ref{theorem::norm_composition} we can compose them to get a reversible 
transducer $\tra''$ uniformizing $\tra$
with $n\cdot 2^{2(n^2+n+1)}\cdot36(n!)^2=2^{O(n^2)}$ states.

\section{Correctness of the construction}
 
First, let us prove that the transducer $\tra'$ is reversible.
 
\begin{lemma}
The \tFT{}  $\tra'$ is reversible.
\end{lemma}
 
\begin{proof}
	To show that $\tra'$ is reversible, we first prove that $\tra'$ is deterministic, and then that it is co-deterministic. This proof is only a case study and does not rely on any new/interesting ideas. The only thing that deserves mentioning is that $\tra$ is weakly branching, that we gave a higher priority to the blocking of the upper component (in order to resolve non-determinism in the case for which both components are blocking), and finally that $\tra$ is co-deterministic.
	
	Let us first show that $\tra'$ is deterministic. Let $a\in\alp$ be a letter, and $s\in\state$ be a state. Four cases depending on the type of $s$:
	\begin{itemize}
		\item $s=\pair{\ov{p}}{\ud{q}}$, the only rules that can be applied are: \fua, \flw, and \fualw\ and they are not compatible since one asks that $p$ has no $a$-successor, the other that $q$ has no successor but $p$ does, and the last that both have an $a$ successor.
		\item $s=\pair{\ud{p}}{\ov{q}}$, this case is symmetrical to the previous case.
		\item $s=\pair{\ov{p}}{\ov{q}}$, the only rules that can be applied are: \buw, \blw, and \bulw\ and they are not compatible since \buw\ asks that the predecessor of $p$ is branching on $a$, \blw\ asks that the predecessor of $q$ is branching on $a$,
		which is not compatible with $\tra$ being weakly branching, and \bulw is not compatible with the others because  both $p$ and $q$ must be the minimal $a$-successor of their predecessor. Moreover there is only one transition per rule since $\tra$ is co-deterministic.
		\item $s=\pair{\ov{p}}{\ov{q}}$, this case is symmetrical to the previous case.
	\end{itemize}
	Thus for every state only one transition is possible. Hence $\tra'$ is deterministic.
	
	We now show that $\tra'$ is co-deterministic. Let $a\in\alp$ be a letter, and $s,s_1,s_2\in\state$ be three states such that $(s_1,a,s)\in\Delta'$ and $(s_2,a,s)\in\Delta'$. We will show that $s_1=s_2$ by a study of four cases depending on the type of $s$:
	\begin{itemize}
		\item $s=\pair{\ov{p}}{\ud{q}}$, then $(s_1,a,s)$ and $(s_2,a,s)$ are of the type \buw, \bla\ or \fualw. Let $p^0$ be the $a$-predecessor of $p$ (which is unique since $\tra$ is co-deterministic) and $q^0$ the predecessor of $q$. Either, $p^0$ is branching on $a$ (which rules out \bla\ since only one state can be branching) and $p=p^0_{\max}$ which rules out \buw. Thus we know that both  $(s_1,a,s)$ and $(s_2,a,s)$ are \fualw\ rules, thus $s_1=s_2=\pair{\ov{p^0}}{\ud{q^0}}$. Otherwise, $p^0$ is branching and $p=p^0_{\min}$. In this case the only possibility for the rules is \buw\ and we obtain $s_1=s_2=\pair{\ud{p^0_{\min}}}{\ud{q}}$. Symmetrically if $q^0$ is branching on $a$, we obtain that $s_1=s_2$. Lastly if none are branching the only possible rule is \fualw, thus $s_1=s_2=\pair{\ov{p^0}}{\ud{q^0}}$.
		\item $s=\pair{\ud{p}}{\ov{q}}$. This case is symmetrical to the previous one with the rules \bua, \blw\ and \fuwla.
		\item $s=\pair{\ud{p}}{\ud{q}}$, then $(s_1,a,s)$ and $(s_2,a,s)$ are of the type \fuw, \flw\ or \bula. If $(s_1,a,s)$ is a rule \fuw\ this means that $p$ has no $a$ successor, thus $(s_2,a,s)$ is also \fuw. Hence $s_1=s_2=\pair{\ov{p}}{\ud{q}}$. If $(s_1,a,s)$ is of type \flw\ this means that $p$ has an $a$ successor but not $q$ thus $(s_2,a,s)$ is also \flw. Hence $s_1=s_2=\pair{\ud{p}}{\ov{q}}$. Finally if they are both \bula, we know that $s_1=s_2=\pair{\ud{p_{\max}}}{\ud{q_{\max}}}$
		\item The last case is symmetrical to the previous one with the rules \fua, \fla\ and \bulw.
	\end{itemize}
	
	This concludes the proof that $\tra'$ is reversible.
 
\end{proof}
 
In order to prove that $\tra$ and $\tra'$ are equivalent, we prove two lemmas that describe the behavior of $\tra'$.
We show that, while going forward, $\tra'$ is always able to chose the smallest branch, and modify its marking.
 
For every word $u \in \vdashv{\alp}^*$, let $\longrun{u} : Q \rightarrow \mathbb{N}$ be the function mapping 
every state $q \in Q$ to the length of the longest run of $\tra$ starting from the configuration $q.u$.
 
\begin{lemma}\label{lemma::seq_state}
Let $u \in \vdashv{\alp}^*$, let $p \neq q$ be two states of $Q$ such that $p$ satisfies $\longrun{u}(p) < |u|$.
\begin{itemize}[noitemsep,nolistsep]
\item
If $\longrun{u}(p) \leq \longrun{u}(q)$, then
$
\live{\rho}_1 : \sta{\live{p}}{\dead{q}} \xrightarrow{u} \sta{\dead{p}}{\dead{q}} \in \Delta'
\ \textup{and} \
\dead{\rho}_1 : \sta{\dead{p}}{\live{q}}  \xrightarrow{u}  \sta{\live{p}}{\live{q}} \in \Delta'.
$
\item
If $\longrun{u}(p) < \longrun{u}(q)$, then
$
\live{\rho}_2 : \sta{\dead{q}}{\live{p}} \xrightarrow{u} \sta{\dead{q}}{\dead{p}} \in \Delta'
\ \textup{and}  \
\dead{\rho}_2 : \sta{\live{q}}{\dead{p}} \xrightarrow{u} \sta{\live{q}}{\live{p}}\in \Delta'.
$
\end{itemize}
Moreover, those four runs produce no outputs.
\end{lemma}
 
\begin{proof}
The proof is by induction on the length of $u$.
If $u = \epsilon$, the result is immediate, since $\longrun{\epsilon}(p) = 0 = |\epsilon|$.
If $u = aw$ for $a \in \vdashv{\alp}$ and $w \in \vdashv{\alp}^*$, suppose that the lemma holds for $w$.

If $p$ has no $a$-successor, $\live{\rho}_1$ and $\dead{\rho}_1$ are runs of $\tra'$, since $\Delta'$ contains the transitions
$(\sta{\live{p}}{\dead{q}},a,\sta{\dead{p}}{\dead{q}})$ and $(\sta{\dead{p}}{\live{q}},a,\sta{\live{p}}{\live{q}})$,
which produce no output.
Moreover, if $\longrun{u}(p) < \longrun{u}(q)$, i.e., $q$ admits an $a$-successor, $\live{\rho}_2$ and $\dead{\rho}_2$ are runs of $\tra'$, since $\Delta'$
contains the transitions
$(\sta{\live{q}}{\dead{p}},a,\sta{\dead{q}}{\dead{p}})$ and $(\sta{\dead{q}}{\live{p}},a,\sta{\live{q}}{\live{p}})$ which,
once again, produce no output.
However, remark that if $\longrun{u}(p) = \longrun{u}(q) = 0$, then neither $\live{\rho}_2$ nor $\dead{\rho}_2$ exists,
since $\tra'$ always checks the continuations of its first component before the second.
 
Now, suppose that $q$ admits an $a$-successor.
Note that we only present the detailed proof of the existence of $\live{\rho}_1$.
The existence of $\live{\rho}_2$ can be proved by swapping the two components of all the states of $\tra'$
and replacing the inequalities with strict inequalities in the following reasoning.
Then, the existence of $\dead{\rho}_1$ and $\dead{\rho}_2$ are derived from the existence of $\live{\rho}_1$ and $\live{\rho}_2$
by substituting, for each state $r \in Q$, $\live{r}$ for $\dead{r}$, $\succmax{r}$ for $\succmin{r}$, and vice versa.

Suppose that $\longrun{u}(p) \leq \longrun{u}(q)$.
Then $q$ admits an $a$-successor.
Let $\succmax{p}$, $\succmin{p}$, $\succmax{q}$, $\succmin{q}$ denote the maximal
and minimal $a$-successors of $p$ and $q$.
Since $\tra$ is weakly branching, $p$ and $q$ admit no other $a$-successor, and $\succmax{p} = \succmin{p}$
or $\succmax{q} = \succmin{q}$.
Moreover, since $\longrun{u}(p) \leq \longrun{u}(q)$,
both $\longrun{u}(\succmax{p})$ and $\longrun{u}(\succmin{p})$ are smaller
than or equal to $\longrun{u}(\succmax{q})$ or $\longrun{u}(\succmin{q})$.
We arrange the different possibilities into three cases, and we expose the existence of
the desired run $\live{\rho}_1$ in each of them, by combining
transitions corresponding to the input letter $a$, and runs resulting from the induction hypothesis.
\begin{enumerate}
\item
If $p$ has a single $a$-successor $\lonesuc{p}$ and $\longrun{u}(\lonesuc{p}) \leq \longrun{u}(\succmin{q})$, then

\begin{tikzpicture}[
                    every edge/.style={draw=black,->,shorten >=0.007pt},
                    every node/.style={anchor=base,text depth=.25ex,text height=1.5ex},
                    text height=1.5ex,text depth=.25ex]

    \node (p1)  at (1.3,0)  {$\live{\rho}_1$ : }; 
    \node (q1)  at (2.2,0)  {$\sta{\live{p}}{\dead{q}}$};       
    \node (q2)  at (4.4,0)  {$\sta{\live{\lonesuc{p}}}{\dead{\succmin{q}}}$};      
    \node (q3)  at (6.6,0)  {$\sta{\dead{\lonesuc{p}}}{\dead{\succmin{q}}}$};     
    \node (q4)  at (8.8,0)  {$\sta{\dead{p}}{\dead{q}}$};      
    \node (q10)  at (10,0)  {$ \in \Delta'.$};

  \draw[->] (q1) edge node[auto] { \scriptsize $a$ } (q2)
            (q2) edge node[auto] {\scriptsize $w$ } (q3)
            (q3) edge node[auto] {\scriptsize $a$ } (q4)
  ;
\end{tikzpicture}
\item
If $p$ has a single $a$-successor $\lonesuc{p}$ and $\longrun{u}(\lonesuc{p}) > \longrun{u}(\succmin{q})$,
then $\longrun{u}(\lonesuc{p}) \leq \longrun{u}(\succmax{q})$, and
%
\tikzset{
  skip loop/.style={to path={-- ++(0,#1) -| (\tikztotarget)}}
}
\begin{tikzpicture}[
                    tip/.style={->,shorten >=0.007pt},every join/.style={rounded corners},
                    every edge/.style={draw=black,->,shorten >=0.007pt},
                    every node/.style={anchor=base,text depth=.25ex,text height=1.5ex},
                    text height=1.5ex,text depth=.25ex]      
                    
      \node (p1)  at (1.3,1)  {$\live{\rho}_1$ : }; 
    \node (q1)  at (2.2,1)  {$\sta{\live{p}}{\dead{q}}$};       
    \node (q2)  at (4.4,1)  {$\sta{\live{\lonesuc{p}}}{\dead{\succmin{q}}}$};      
    \node (q3)  at (6.6,1)  {$\sta{\live{\lonesuc{p}}}{\live{\succmin{q}}}$};     
    \node (q4)  at (8.8,1)  {$\sta{\live{\lonesuc{p}}}{\dead{\succmax{q}}}$};     
 
    \node (q6)  at (2.2,0)  {$\sta{\dead{\lonesuc{p}}}{\dead{\succmax{q}}}$};       
    \node (q7)  at (4.4,0)  {$\sta{\dead{\lonesuc{p}}}{\live{\succmin{q}}}$};      
    \node (q8)  at (6.6,0)  {$\sta{\dead{\lonesuc{p}}}{\dead{\succmin{q}}}$};     
    \node (q9)  at (8.8,0)  {$\sta{\dead{p}}{\dead{q}}$};      
    \node (q10)  at (10,0)  {$ \in \Delta'.$};      
    
 \node at (5.5,0.7) {\scriptsize{$w$}}; 

  { [start chain]
    \chainin (q4);
    \chainin (q6)    [join=by {skip loop=-5mm,tip}];
  }
    \draw[->] (q1) edge node[auto] { \scriptsize $a$ } (q2)
            (q2) edge node[auto] {\scriptsize $w$ } (q3)
            (q3) edge node[auto] {\scriptsize $a$ } (q4)
  ;
    \draw[->] (q6) edge node[auto] { \scriptsize $a$ } (q7)
            (q7) edge node[auto] {\scriptsize $w$ } (q8)
            (q8) edge node[auto] {\scriptsize $a$ } (q9)
  ;
\end{tikzpicture}
\item
If $\succmax{p} \neq \succmin{p}$, then $q$ has a single $a$-successor $\lonesuc{q}$ and
 $\longrun{u}(\succmax{p}),\longrun{u}(\succmin{p}) \leq \longrun{u}(\lonesuc{q})$. Hence
\begin{tikzpicture}[
                    every edge/.style={draw=black,->,shorten >=0.007pt},
                    every node/.style={anchor=base,text depth=.25ex,text height=1.5ex},
                    text height=1.5ex,text depth=.25ex]

    \node (p1)  at (1.3,0)  {$\live{\rho}_1$ : }; 
    \node (q1)  at (2.1,0)  {$\sta{\live{p}}{\dead{q}}$};       
    \node (q2)  at (4.2,0)  {$\sta{\live{\succmax{p}}}{\dead{\lonesuc{q}}}$};      
    \node (q3)  at (6.3,0)  {$\sta{\dead{\succmax{p}}}{\dead{\lonesuc{q}}}$};     
    \node (q4)  at (8.4,0)  {$\sta{\live{\succmin{p}}}{\dead{\lonesuc{q}}}$};      
    \node (q5)  at (10.5,0)  {$\sta{\dead{\succmin{p}}}{\dead{\lonesuc{q}}}$};      
    \node (q6)  at (12.6,0)  {$\sta{\dead{p}}{\dead{q}}$};      
    \node (q10)  at (13.7,0)  {$ \in \Delta'.$};

  \draw[->] (q1) edge node[auto] { \scriptsize $a$ } (q2)
            (q2) edge node[auto] {\scriptsize $w$ } (q3)
            (q3) edge node[auto] {\scriptsize $a$ } (q4)
            (q4) edge node[auto] {\scriptsize $w$ } (q5)
            (q5) edge node[auto] {\scriptsize $a$ } (q6)
  ;
\end{tikzpicture}
\end{enumerate}
Finally, since $\tra$ is co-deterministic and $p \neq q$, no $a$-successor of $p$ matches any $a$-successor of $q$, and, by
the definition of $\mu'$ and the induction hypothesis, the run $\live{\rho}_1$ produces no output in each of the three cases.
\end{proof}

\begin{lemma}\label{lemma::acc_state}
Let $u \in \vdashv{\alp}^*$, $p \in Q$ and let $u'$ be the longest prefix of $u$ such that $p$ has a $u'$-successor.
Then there exists a run $\rho = (p,u',q)$ of $\tra$ such that 
$|u'| = \longrun{u}(p)$ and $\rho' = (\sta{\live{p}}{\dead{p}},u',\sta{\live{q}}{\dead{q}}) \in \Delta'$.
Moreover, $\mu'(\rho') = \mu(\rho)$. 
\end{lemma}
 
\begin{proof}
The proof is by induction on the length of $u$.
If $u = \epsilon$, the lemma holds immediately by choosing the empty run $(p,\epsilon,p)$.
Now suppose that $u = aw$ for some $a \in \vdashv{\alp}$, $w \in \vdashv{\alp}^*$,
and let us suppose that the lemma holds for $w$.
Let $\succmax{p}$ and $\succmin{p}$ denote the maximal and minimal $a$-successors of $p$.
Since $\tra$ is weakly branching, $p$ admits no other $a$-successor.
Therefore, $\longrun{w}(\succmax{p}) = \longrun{u}(p) -1$ or $\longrun{w}(\succmin{p}) = \longrun{u}(p) -1$.
We consider three cases, and we expose the run $\rho'$ in each of them by combining transitions corresponding to $a$, the induction hypothesis, and  the runs from Lemma \ref{lemma::seq_state}.
\begin{itemize}
\item
If $p$ has a single $a$-successor $p_0$, then $\longrun{}(p_0) = \longrun{}(p) -1$ and
by the induction hypothesis there is a run $(p_0,w',q) \in \Delta$ satisfying the lemma for $w$.
Let $v_a = \mu(p,a,p_0)$ and $v_w = \mu(p_0,w',q)$.
Then

\begin{tikzpicture}[
                    every edge/.style={draw=black,->,shorten >=0.007pt},
                    every node/.style={anchor=base,text depth=.25ex,text height=1.5ex},
                    text height=1.5ex,text depth=.25ex]

    \node (p1)  at (1.6,0)  {$\rho'$ : }; 
    \node (q1)  at (2.4,0)  {$\sta{\live{p}}{\dead{p}}$};       
    \node (q2)  at (4.6,0)  {$\sta{\live{p_0}}{\dead{p_0}}$};   
    \node (q3)  at (7.1,0)  {$\sta{\live{q}}{\dead{q}}$};      
    \node (q10)  at (8.7,0)  {$ \in \Delta'.$};

  \draw[->] (q1) edge node[auto] { \scriptsize $a|v_a$ } (q2)
            (q2) edge node[auto] {\scriptsize $w|v_w$ } (q3)
  ;
\end{tikzpicture}
\item
If $\succmax{p}$ and $\succmin{p}$ are distinct and $\longrun{}(\succmin{p}) < \longrun{}(\succmax{p}) = \longrun{}(p) -1$,
by the induction hypothesis there is a run $(\succmax{p},w',q) \in \Delta$ satisfying the lemma for $w$.
Let $v_a = \mu(p,a,\succmax{p})$ and $v_w = \mu(\succmax{p},w',q)$.
Then,

\begin{tikzpicture}[
                    every edge/.style={draw=black,->,shorten >=0.007pt},
                    every node/.style={anchor=base,text depth=.25ex,text height=1.5ex},
                    text height=1.5ex,text depth=.25ex]

    \node (p1)  at (1.6,0)  {$\live{\rho}_1$ : }; 
    \node (q1)  at (2.4,0)  {$\sta{\live{p}}{\dead{p}}$};       
    \node (q2)  at (4.6,0)  {$\sta{\live{\succmax{p}}}{\dead{\succmin{p}}}$};      
    \node (q3)  at (7.1,0)  {$\sta{\live{\succmax{p}}}{\live{\succmin{p}}}$};     
    \node (q4)  at (9.6,0)  {$\sta{\live{\succmax{p}}}{\dead{\succmax{p}}}$};      
    \node (q5)  at (12.1,0)  {$ \sta{\live{q}}{\dead{q}}$};
    \node (q10)  at (13.7,0)  {$ \in \Delta'.$};

  \draw[->] (q1) edge node[auto] { \scriptsize $a|\epsilon$ } (q2)
            (q2) edge node[auto] {\scriptsize $w'|\epsilon$ } (q3)
            (q3) edge node[auto] {\scriptsize $a|v_a$ } (q4)
            (q4) edge node[auto] {\scriptsize $w'|v_w$ } (q5)
  ;
\end{tikzpicture}
\item
If $\succmax{p}$ and $\succmin{p}$ are distinct and $\longrun{}(\succmax{p}) \leq \longrun{}(\succmin{p}) = \longrun{}(p) -1$,
by the induction hypothesis there is a run $(\succmin{p},w',q) \in \Delta$ satisfying the lemma for $w$.
Let $v_a = \mu(p,a,\succmin{p})$ and $v_w = \mu(\succmin{p},w',q)$.
Then

\begin{tikzpicture}[
                    every edge/.style={draw=black,->,shorten >=0.007pt},
                    every node/.style={anchor=base,text depth=.25ex,text height=1.5ex},
                    text height=1.5ex,text depth=.25ex]

    \node (p1)  at (1.6,0)  {$\live{\rho}_1$ : }; 
    \node (q1)  at (2.4,0)  {$\sta{\live{p}}{\dead{p}}$};       
    \node (q2)  at (4.6,0)  {$\sta{\live{\succmax{p}}}{\dead{\succmin{p}}}$};      
    \node (q3)  at (7.1,0)  {$\sta{\dead{\succmax{p}}}{\dead{\succmin{p}}}$};     
    \node (q4)  at (9.6,0)  {$\sta{\live{\succmin{p}}}{\dead{\succmin{p}}}$};      
    \node (q5)  at (12.1,0)  {$ \sta{\live{q}}{\dead{q}}$};
    \node (q10)  at (13.7,0)  {$ \in \Delta'.$};

  \draw[->] (q1) edge node[auto] { \scriptsize $a|\epsilon$ } (q2)
            (q2) edge node[auto] {\scriptsize $w'|\epsilon$ } (q3)
            (q3) edge node[auto] {\scriptsize $a|v_a$ } (q4)
            (q4) edge node[auto] {\scriptsize $w'|v_w$ } (q5)
  ;
\end{tikzpicture}
\qedhere
\end{itemize}
\end{proof}
 
\begin{corollary}\label{corollary::equiv}
The \tFT{} $\tra'$ is equivalent to $\tra$.
\end{corollary}
 
\begin{proof}
Let $u \in \alp^*$, and let $\vdashv{u}$ denote $\vdash u \dashv$.
If $u \notin \lang_{\tra}$, there exists no accepting run of $\tra$ on $\vdashv{u}$.
Therefore $\longrun{\vdashv{u}}(q_I) < |\vdashv{u}|$, and by Lemma \ref{lemma::acc_state},
the run of $\tra'$ on $\vdashv{u}$ starting from the initial configuration
$\sta{\live{q_I}}{\dead{q_I}}.\vdashv{u}$ will eventually reach the configuration
$\vdash u'.\sta{\live{q}}{\dead{q}}.u'' \dashv$, such that $q$ admits no $a$-successor,
where $a$ denotes the first letter of $u''\dashv$.
Note that this configuration is rejecting, since according to the transition relation $\Delta'$ of $\tra'$, the only
candidate to be an $a$-successor of $\sta{\live{q}}{\dead{q}}$ is the pair $\sta{\dead{q}}{\dead{q}}$,
which is not part of $Q'$ by definition.

Conversely, if $u \in \lang_{\tra}$, there exists an accepting run $\rho$ of $\tra$ on $\vdashv{u}$.
Therefore $\longrun{\vdashv{u}}(q_I) = |\vdashv{u}|$, and by Lemma \ref{lemma::acc_state}, the run $\rho'$ of $\tra'$ on $\vdashv{u}$ starting from the initial configuration
$\sta{\live{q_I}}{\dead{q_I}}.\vdashv{u}$ will eventually end in the configuration $\vdashv{u}.\sta{\live{q_F}}{\dead{q_F}}$,
since by convention $q_F$ is the only possible target of a transition of the form $(q,\dashv,q_F)$.
Moreover, $\rho'$ produces the same output as $\rho$.
This concludes the proof.
\end{proof}


\end{document}